\documentclass[12pt,tightenlines]{revtex4}
\usepackage{amsfonts}
\usepackage{amsmath}
\usepackage{txfonts}
\usepackage{graphicx}
\usepackage{amssymb}
\usepackage{color}
\usepackage{wrapfig,lipsum,booktabs}
\usepackage{floatrow}
\usepackage{epstopdf}
\usepackage{hyperref}
\usepackage[normalem]{ulem}

\newcommand{\beginsupplement}{
        \setcounter{table}{0}
        \renewcommand{\thetable}{S\arabic{table}}
        \setcounter{figure}{0}
        \renewcommand{\thefigure}{S\arabic{figure}}
     }

\newcommand{\bdry}{\partial}
\newcommand{\Zgr}{{\sf Z}}
\newcommand{\Bgr}{{\sf B}}
\newcommand{\Hgr}{{\sf H}}
\newcommand{\im}{\operatorname{im}}
\newcommand{\Betti}{\beta}
\newcommand{\ot}{\leftarrow}
     
\begin{document}
\title{Robust spatial memory maps encoded in networks with transient connections}
\author{Andrey Babichev$^{1}$, Dmitriy Morozov$^{2}$ and Yuri Dabaghian$^{1}$\textsuperscript{*},}
\affiliation{$^1$Department of Computational and Applied Mathematics, Rice University, Houston, TX 77005, USA \\
$^2$Lawrence Berkeley National Laboratory, Berkeley, CA 94720, USA\\
$^{*}$e-mail: dabaghian@gmail.com}
\date{\today}

\begin{abstract} 
\vspace{10 mm}

\textbf{Abstract}. The spiking activity of principal cells in mammalian hippocampus encodes an internalized 
neuronal representation of the ambient space---a cognitive map. Once learned, such a map enables the animal 
to navigate a given environment for a long period. However, the neuronal substrate that produces this map remains 
transient: the synaptic connections in the hippocampus and in the downstream neuronal networks never cease to 
form and to deteriorate at a rapid rate. How can the brain maintain a robust, reliable representation of space using 
a network that constantly changes its architecture? Here, we demonstrate, using novel Algebraic Topology techniques, 
that cognitive map’s stability is a generic, emergent phenomenon. The model allows evaluating the effect produced by 
specific physiological parameters, e.g., the distribution of connections' decay times, on the properties of the cognitive 
map as a whole. It also points out that spatial memory deterioration caused by weakening or excessive loss of the 
synaptic connections may be compensated by simulating the neuronal activity. Lastly, the model explicates functional 
importance of the complementary learning systems for processing spatial information at different levels of spatiotemporal 
granularity, by establishing three complementary timescales at which spatial information unfolds. Thus, the model 
provides a principal insight into how can the brain develop a reliable representation of the world, learn and retain 
memories despite complex plasticity of the underlying networks and allows studying how instabilities and memory 
deterioration mechanisms may affect learning process.

\vspace{10 mm}

\textbf{Significance Statement}. We explain how reliable representations of the world can emerge in networks with 
transient synaptic architectures. We study properties of the hippocampal cognitive map produced by place cell assemblies 
that recycle at the working memory timescale and demonstrate that the resulting ``transient'' network can represent 
the topology of the environment. We show that 1) this is a generic phenomenon, implementable via different mechanisms; 
2) that deterioration of the memory map caused by excessive loss of the synaptic connections may be compensated by 
simulating the neuronal activity in the hippocampal network; 3) evaluate the effect produced by specific physiological 
parameters, e.g., the distribution of connections' decay times; 4) explicate three complementary timescales at which 
spatial information is processed in the brain.

\end{abstract}

\maketitle

\newpage

\section{Introduction}
\label{section:intro}

Functioning of the biological networks relies on synaptic and structural plasticity processes taking place at various 
spatiotemporal timescales \cite{Bi, Leuner,Caroni}. For example, the so-called place cells in mammalian hippocampus 
learn to spike within specific locations of a new environment (their respective place fields) in a matter of minutes and 
then exhibit slow tuning of their firing rates for weeks  \cite{Best,Frank,Karlsson}. The synaptic architecture of the 
hippocampal network constantly changes due to formation, adaptation and pruning of the synaptic connections via fast 
and slow plasticity mechanisms. In particular, detailed analyses of spike time statistics suggest that the place cells group 
into transient ``assemblies'' that may appear and disappear at working and intermediate memory timescales 
\cite{Harris,Buzsaki1}.

The fact that the hippocampal network has a dynamic synaptic architecture poses a principal question: how can a 
rapidly rewiring network produce and sustain a stable cognitive map? How can it provide the downstream networks 
with stable spatial information? In the following, we address this question by modeling a population of dynamical 
place cell assemblies and studying the effect produced by the network's transience on the large-scale representation 
of space, using algebraic topology tools. In particular, we demonstrate that despite rapid changes in its synaptic 
architecture, a transient cell assembly network can encode a stable large-scale topological map within a biologically 
plausible period.

\section{The topological model}
\label{section:model}

\textbf{General outline}. Our model of the hippocampal network is based on a schematic representation of the 
information provided by a population of spiking place cells in a given environment \cite{Dabaghian,Arai, Basso,Babichev1}. 
First, a group of coactive place cells, $c_0$, $c_1$, \ldots, $c_n$ is represented by an abstract simplex $\sigma = [c_0, c_1, 
\ldots,c_n]$---a basic object from algebraic topology that may be viewed geometrically as a $n$-dimensional tetrahedron 
with $n + 1$ vertexes (see Methods). Due to spatial tuning of the place cell activity, each individual coactivity simplex may 
also be viewed as a representation of the spatial overlap between the corresponding place fields. Together, the full collection 
of such simplexes forms a simplicial ``coactivity'' complex $\mathcal{T}$ that represents spatial connectivity among the 
place fields that cover the environment $\mathcal{E}$---the place field map $M_\mathcal{E}$ (see Methods). 

\begin{figure}[!h]
\includegraphics[scale=0.83]{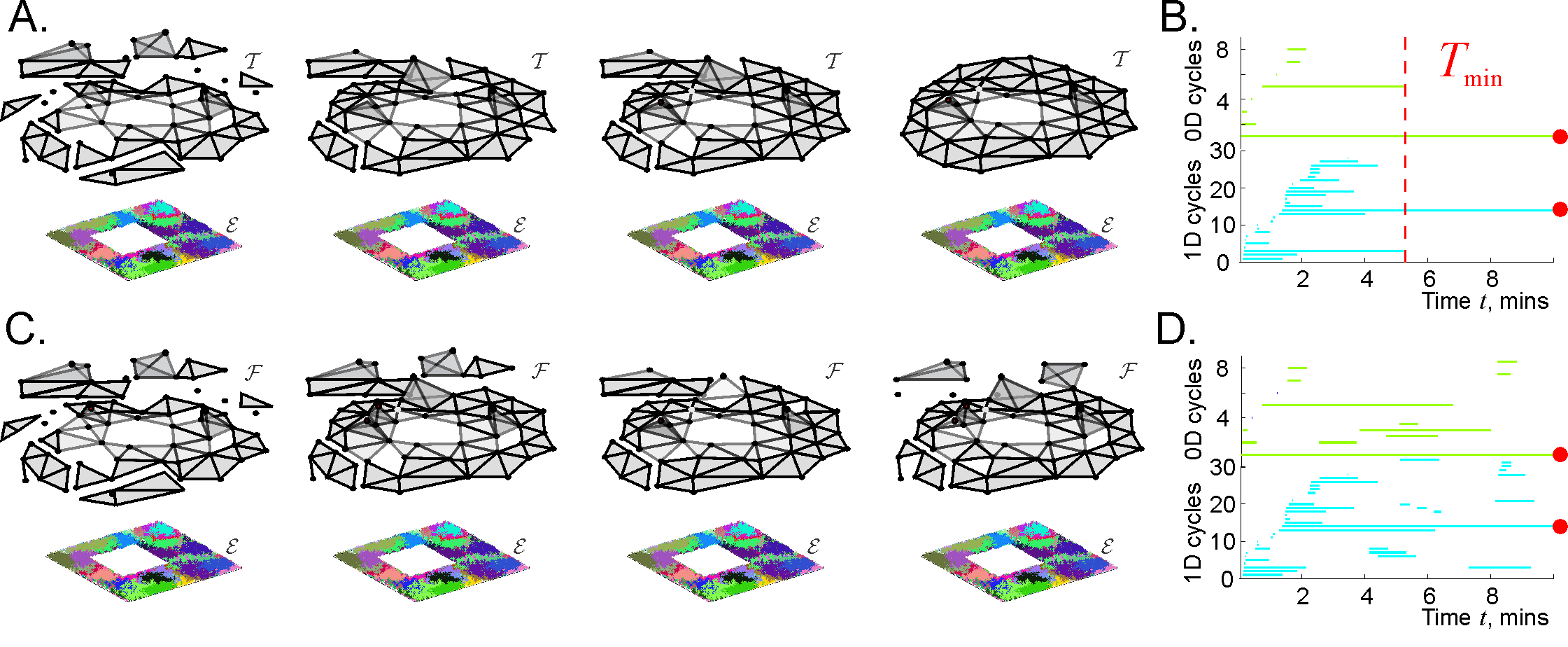}
\caption{{\footnotesize\textbf{Topological structure of the perennial and decaying coactivity complexes.}
A: Simulated place field map $M_\mathcal{E}$ of a small planar environment $\mathcal{E}$ with a square hole in the middle
(see Methods). Four consecutive snapshots illustrate the temporal dynamics of the coactivity complex: at the early stages of 
navigation the complex is small and fragmented, but as the topological information accumulates, the transient topological loops 
disappear, yielding a stable topological shape that is equivalent to the shape of the underlying environment. B: The timelines 
of topological loops in a steadily growing simplicial complex computed using persistent homology methods: the timelines of 
disconnected pieces ($0D$ loops) are shown by light-blue lines and the timelines of one-dimensional holes ($1D$ loops) are 
light-green. Most loops are spurious, i.e., correspond to accidental, short-lasting structures in $\mathcal{T}(M_\mathcal{E})$. 
The persistent topological loops (marked by red dots) represent physical features of the environment $\mathcal{E}$, i.e., its 
main connected component and the central hole. The time $T_{\min}$ required to eliminate the spurious loops can serve as 
a theoretical estimate of the minimal time needed to learn path connectivity of the environment. C: If the simplexes may not 
only appear but also disappear, then the structure of the resulting ``flickering'' coactivity complex $\mathcal{F}(M_\mathcal{E})$ 
may never saturate. D: The timelines of the topological loops in such complex may remain interrupted by opening and closing 
topological gaps produced by decays and reinstatements of its simplexes.}}
\label{Figure1}
\end{figure}

The formation of the coactivity complex represents the process of accumulating the topological information supplied by the 
place cell activity. At the beginning of navigation, when a few coactive place cells had time to fire, the complex 
$\mathcal{T}(M_\mathcal{E})$ contains a few simplexes that (typically) form several disconnected agglomerates 
(subcomplexes of $\mathcal{T}$), riddled with holes. Biologically, those may be viewed as fragments and gaps of the 
emerging cognitive map (Fig.~\ref{Figure1}A). If the parameters of spiking activity fall within the biological range of values, 
then, as more and more instances of coactivity are produced, the coactivity complex $\mathcal{T}(M_\mathcal{E})$ grows 
and eventually saturates, assuming a shape that is topologically equivalent to the shape of the navigated environment. 

Mathematically, the topological structure of a steadily growing coactivity complex can be described using persistent 
homology theory methods \cite{Ghrist,Zomorodian}. In particular, this theory allows detecting topological loops in 
$\mathcal{T}(M_\mathcal{E})$ (i.e., closed chains of simplexes identified up to topological equivalence \cite{Friedman}), 
on a moment-by-moment basis (Fig.~\ref{Figure1}B). Such loops provide a convenient semantics for describing how the shape 
of $\mathcal{T}(M_\mathcal{E})$ unfolds in time. For example, the number of inequivalent topological loops that can be 
contracted to a zero-dimensional vertex defines the number of the connected components in $\mathcal{T}(M_\mathcal{E})$; 
the number of loops that contract to a one-dimensional chain of links defines the number of holes and so forth.  
In the literature (see, e.g., \cite{Hatcher}), the number of $k$-dimensional topological loops in a space $X$ is referred to as its 
$k$-th Betti number, $b_k(X)$, and the list of all Betti numbers defines the topological barcode, 
$\mathfrak{b}(X) = (b_0(X),b_1(X),\ldots)$, that specifies the topological shape of $X$. For example, the 
simply connected, square environment $\mathcal{E}$ with a single hole in the middle (Fig.~\ref{Figure1}A and Methods) has 
the Betti numbers $b_0(\mathcal{E}) = b_1(\mathcal{E}) = 1$, $b_{k >1}(\mathcal{E}) = 0$, and hence its topological 
barcode is $\mathfrak{b}(\mathcal{E}) = (1,1,0,0,\ldots)$.

The construction of the coactivity complexes may be adopted to reflect physiological aspects of the hippocampal network. 
For example, the simplexes of the coactivity complex may represent not just arbitrary combinations of coactive cells, but 
the actual, physiological neuronal assemblies---groups of cells that jointly elicit spiking activity in the downstream neurons. 
As mentioned in the introduction, these assemblies are unstable, transient structures that are recycled, according to 
different estimates, at the timescale between minutes to hundreds of milliseconds \cite{Harris, Buzsaki1}. In order to 
represent this transience, the coactivity complex has to acquire a qualitatively different dynamics: its simplexes should 
be allowed to appear and to disappear, i.e., ``flicker,'' following the appearances and disappearances of the corresponding 
cell assemblies. As a result, certain parts of the coactivity complex may inflate, while others may shrink, at different rates 
and in various sequences (Fig.~\ref{Figure1}C). The topological structure of such a ``flickering'' coactivity complex
 $\mathcal{F}(M_\mathcal{E})$ cannot, in general, be described using ordinary persistent homology theory methods 
 (Fig.~\ref{Figure1}D), and requires a different mathematical apparatus---Zigzag persistent homology theory, outlined in the 
 Methods section and in \cite{Carlsson1,Carlsson2, Edelsbrunner}.

\textbf{Implementation}. An efficient implementation of the coactivity complex is based on a classical ``cognitive graph'' 
model of the hippocampal network \cite{Burgess,Muller, Babichev2}. In this model, each active place cell $c_i$ corresponds 
to a vertex $v_i$ of graph $\mathcal{G}$, and the connections between pairs of cells (physiological or functional) are 
represented by the links $\varsigma^2_{ij} = [v_i, v_j]$ of $\mathcal{G}$. The assemblies of place cells $c_1$, $c_2$, ..., $c_n$ 
(``synaptically interconnected networks'' in terminology of \cite{Buzsaki1}) can then be naturally interpreted as fully 
interconnected subgraphs between the corresponding vertexes, i.e., as the maximal cliques $\varsigma = [v_1, v_2, \ldots, v_n]$ 
of $\mathcal{G}$ \cite{ Babichev1, Hoffman}. The connection with the topological model described above comes from the 
observation that cliques, as combinatorial objects, can be viewed as simplexes spanned by the same sets of vertexes. 
In other words, the collection of cliques of any graph $G$ defines the so-called clique complex $\Sigma(G)$ \cite{Jonsson}, 
and hence the set of the coactivity cliques of $\mathcal{G}$ can be viewed as the coactivity complex associated with the 
cognitive graph model. Such a complex effectively accumulates the information about place cell coactivity at various timescales, 
capturing the correct topology of planar \cite{Dabaghian,Arai,Basso, Babichev1} and voluminous \cite{Hoffman} environments 
within minutes, which provides a suitable ground for constructing a ``flickering complex'' model for the network of dynamical 
cell assemblies. 
Specifically, one can use a coactivity graph $\mathcal{G}$ with appearing and disappearing (flickering) links, and evaluate 
the topological shape of the corresponding flickering coactivity complex, using Zigzag persistent homology theory techniques. 
This constitutes a simple phenomenological model that connects the information provided by individual dynamical place cell 
assemblies and their physiological properties (e.g., the rate of their transience) to the structure of the large-scale 
topological maps encoded by the cell assembly network as a whole. 

The model discussed below is implemented under the following principal assumptions.

I. \emph{Decay of the synaptic connections}. A simple description of a transient network can be given in terms of the 
probabilities of the links' appearances and disappearances at a given moment. For the latter, we adopt a basic ``decay'' 
model, in which an existing link $\varsigma^2_{ij}$ between cells $c_i$ and $c_j$ can disappear with the probability 
$$p_{ij}(t) = \frac{1}{\tau_{ij}}e^{-t/\tau_{ij}},$$ where the time $t$ is counted from the moment of the link's last 
appearance and the parameter $\tau_{ij}$ defines its mean decay time. The decay times of the higher order cliques in 
the coactivity graph (i.e., of the higher order cell assemblies in the hippocampal network) are then defined by the 
corresponding link's half-lives.

In a physiological cell assembly network, the decay times $\tau_{ij}$ should be distributed around a certain mean 
$\tau$ with a certain statistical variance \cite{ Sayer}. However, in order to simplify the current model and to facilitate 
the interpretation of its outcomes, we attribute a single value $\tau_{ij} = \tau$ to all links in $\mathcal{G}$ and use a 
unified distribution
\begin{equation}
p_0(t) = \frac{1}{\tau}e^{-t/\tau},
\label{decay}
\end{equation}
to describe the deterioration of all synaptic connections within all cell assemblies. Thus, $\tau$ will be the only parameter 
that describes the decay of the synaptic connections in the model. We will therefore use the notations $\mathcal{G}_{\tau}$ 
and $\mathcal{F}_{\tau}$ to refer, respectively, to the flickering coactivity graph with decaying connections and to the resulting 
flickering clique coactivity complex with decaying simplexes. 

II. \emph{Appearances and rejuvenations of the synaptic connections}. We will assume that a connection $\varsigma^2_{ij}$ 
in the graph $\mathcal{G}$ appears if the cells $c_i$ and $c_j$ become active within a $w = 1/4$ second period (biologically, 
this corresponds to consecutive periods of the $\theta$-rhythm \cite{Mizuseki,Arai}). The subsequent coactivities of the pair 
$[c_i, c_j]$ either reinstate the link $\varsigma^2_{ij}$ (if it has disappeared by that moment) or rejuvenates it (i.e., its decay 
restarts). 
As a result, the links' actual or \emph{effective} mean lifetime $\tau_{e}$ may differ from the proper decay time $\tau$, which 
defines the expected lifetime of an unperturbed connection. Indeed, if the connection $\varsigma^2_{ij}$ that appeared at a 
moment $t_1$ does not disappear by the moment $t_2$ when the pair of cells $[c_i, c_j]$ reactivates, then its expected lifetime 
becomes $t_2 - t_1 + \tau$. If it does not decay before being ``rejuvenated'' again at a later moment $t_3$, then its expected 
lifetime is $t_3 - t_1 + \tau$ and so forth. Notice however, that since place cells' spiking in learned environments is stable 
\cite{Thompson}, the vertexes in the coactivity complex $\mathcal{F}_{\tau}$ appear with the first activation of the corresponding 
place cells and then never disappear. 

III. \emph{Fixed geometric parameters}. The series of instances at which a given combination of cells may become active is 
defined by the geometry of the place field map $M_\mathcal{E}$ and by the times of the rat's visits into the locations where 
the corresponding place fields overlap \cite{McNaughton,Shapiro}. In order to focus on the dependence of the topology of 
$\mathcal{F}_{\tau}(M_\mathcal{E})$ on the links' decay time, we selected a specific trajectory $\gamma(t)$ and a particular 
place field map $M_\mathcal{E}$ that induces a coactivity complex with correct topology in the ``perennial'' ($\tau = \infty$) 
limit \cite{ Dabaghian,Arai}, and studied how the dynamics of the Betti numbers $b_k(\mathcal{F}_{\tau}(M_\mathcal{E}))$ 
depends on $\tau$. In the following, we will therefore omit references to the place field map in the notations of the coactivity 
graph or the coactivity complex, and write simply $\mathcal{G}_{\tau}$ and $\mathcal{F}_{\tau}$.

IV. \emph{Restricted dimensionality}. Lastly, we note that although the coactivity complex is multidimensional \cite{ Babichev1}, 
for a topological description of a planar environment it suffices to consider only the two-dimensional skeleton of $\mathcal{F}_{\tau}$, 
i.e., the collection of second ($\varsigma^2$) and third ($\varsigma^3$) order connections (i.e., second or third order cliques of 
$\mathcal{G}_{\tau}$ or two- or three-vertex simplexes of $\mathcal{F}_{\tau}$). Thus, in the following we will compute the 
coactive pairs and triples of the simulated neurons in order to study the topological properties of $\mathcal{F}_{\tau}$ as function 
of $\tau$. 

A priori, one would expect that if $\tau$ is too small, then the flickering complex $\mathcal{F}_{\tau}$ deteriorates too rapidly 
to produce a stable topological representation of the environment. In contrast, if $\tau$ is too large, then the effect of synaptic 
deterioration will not be significant. Thus, our goal will be to identify just how rapidly the coactivity simplexes can recycle while 
preserving the net topological structure of $\mathcal{F}_{\tau}$. Physiologically, this will define how rapidly the hippocampal 
cell assemblies can rewire without jeopardizing the integrity of the topological map of the environment.

\section{Results}
\label{section:results}

\begin{figure}[!h]
\includegraphics[scale=0.83]{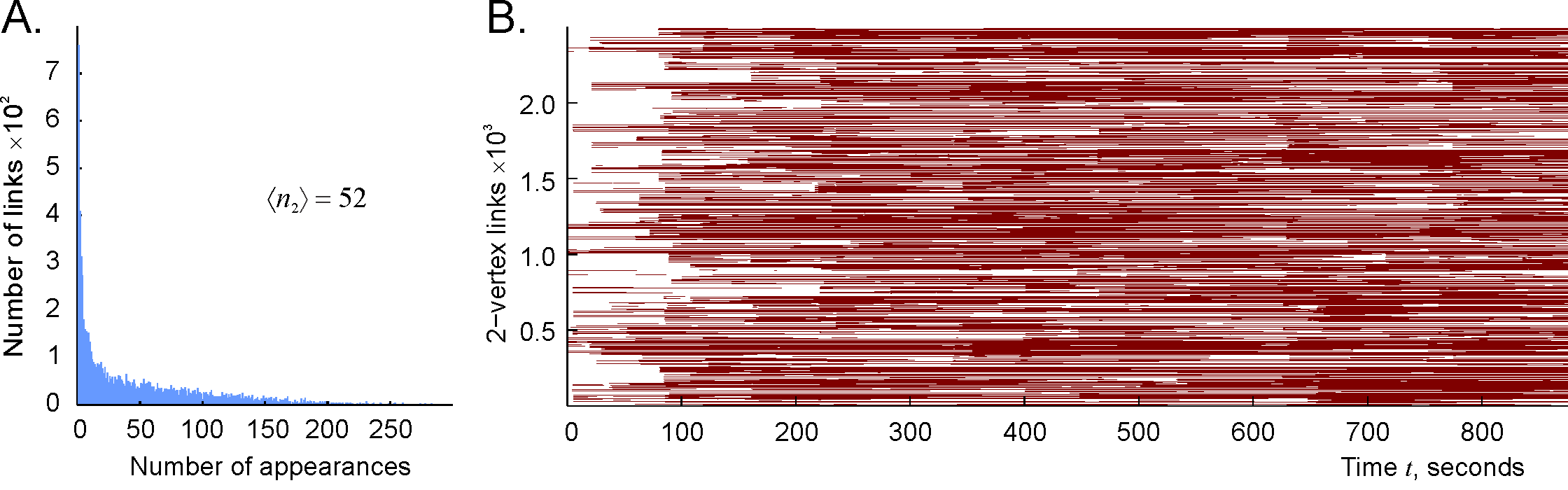}
\caption{{\footnotesize\textbf{Time course of pairwise coactivity.} A: Histogram of the number of times that a given connection 
$\varsigma^2$ activates, computed for the place field map illustrated on Fig.~\ref{Figure1} (for the corresponding occupancy map, 
see Methods). 
On average, a connection is activated about $50$ times over the $25$ min navigation period, i.e., once every $30$ secs, although 
most links activate only a few times and some of them appear hundreds of times. B: Timelines of the links in the coactivity graph.}}
\label{Figure2}
\end{figure}

To start the simulations, we reasoned that in order for the flickering complex $\mathcal{F}_{\tau}$ to accumulate a 
sufficient number of simplexes and to capture the topology of the environment, its simplexes should not disappear 
between two consecutive coactivities of the corresponding cell groups. In other words, the characteristic lifetime of the 
links of the coactivity graph should exceed the typical interval between two consecutive activations of the corresponding 
cell pairs. In the simulated map, a typical link $\varsigma^2$ in the connectivity graph $\mathcal{G}$ activates about 
$\langle n_2 \rangle = 50$ times during the $T_{tot} = 25$ min navigation period, i.e., the mean activation frequency 
is $f_2 \approx 1/30$ Hz (Fig.~\ref{Figure2}). Hence, in order to make room for the rejuvenation effects, we first tested 
the decay time of $\tau = 100$ secs, which is about three times longer than the inter-activity period and by an order of 
magnitude smaller than the total navigation time $\tau \approx T_{tot}/15$. 

\begin{figure}[!h]
\includegraphics[scale=0.83]{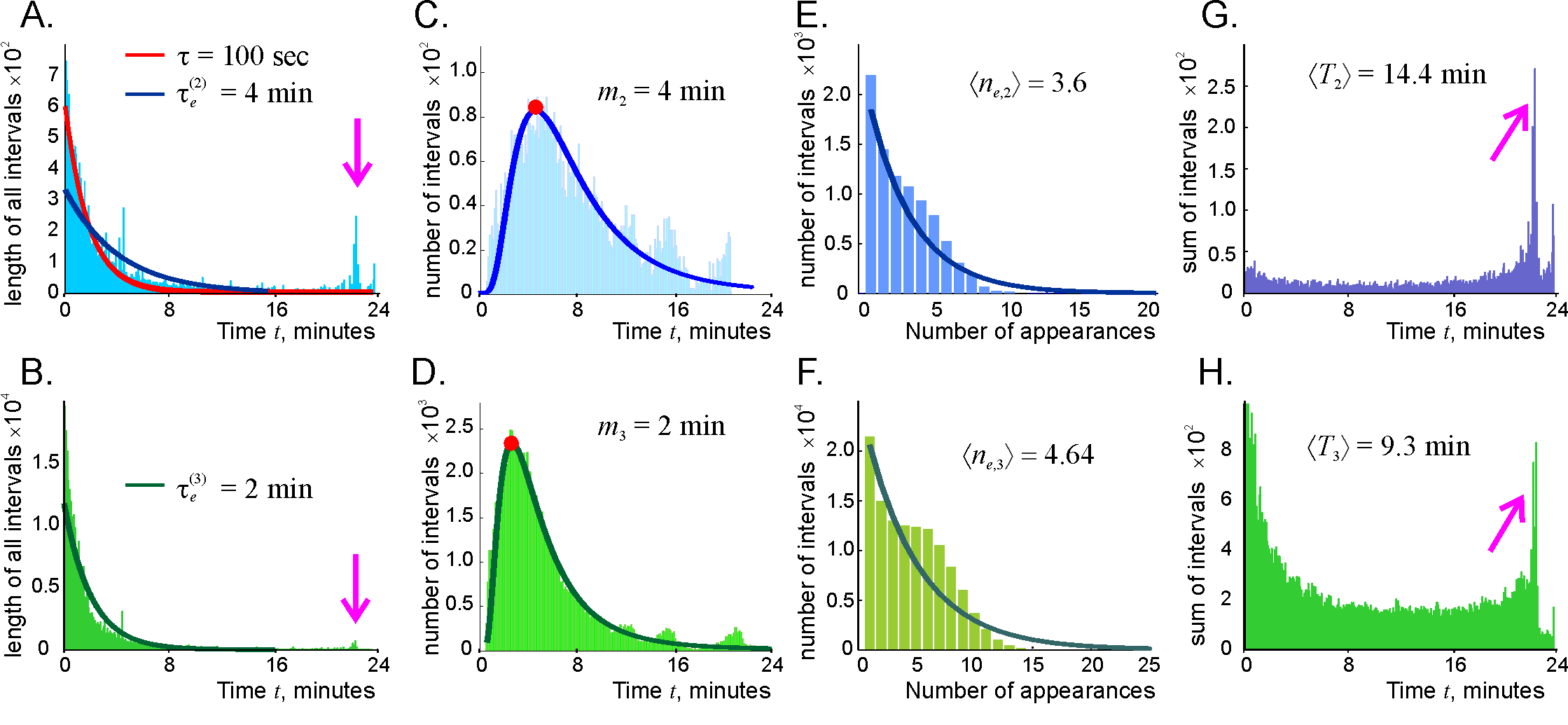}
\caption{{\footnotesize\textbf{Connection dynamics.} A: The histogram of the time intervals between the consecutive births and deaths 
of the links, $\Delta t_{\varsigma_2,i} = t^{(d_i)}_{\varsigma_2} - t^{(b_i)}_{\varsigma_2}$, where the index $i$ 
enumerates the birth ($b_i$) and the death ($d_i$) events. The mean of the exponential that fits the left side of the 
histogram (dark-blue line) is shown at the top of the panel. The pink arrow points at the population of the ``survivor'' links. 
The red line marks the distribution (\ref{decay}). 
B: Similar histogram for the third order simplexes. The histograms of the lifetimes averaged over all the instances of a 
given simplexes' appearances $\Delta t_{\varsigma_2} = \langle t^{(d_i)}_{\varsigma_2} - t^{(b_i)}_{\varsigma_2}\rangle_i$ 
(panel C) and $\Delta t_{\varsigma_3} = \langle t^{(d_i)}_{\varsigma_3} - t^{(b_i)}_{\varsigma_3}\rangle_i$ (panel D). 
The modes of the resulting lognormal distributions (solid lines), $m_2$ and $m_3$, correspond to the means shown on 
panels A and B. The histograms of the number of times the link and the triple connections activate during the navigation 
period are shown on the panels E and F. The exponential fits are shown by solid lines, with the means shown at the top 
of the panels. The distributions of total existence times for the second (panel G) and third (panel H) order simplexes, with 
the averages that are approximately equal to the product of the mean effective lifetime and the mean number of appearances, 
$\Delta T_{e,\varsigma}\approx n_{e,\varsigma}\tau_{e,\varsigma}$.}}
\label{Figure3}
\end{figure}

The histogram of the time intervals between the consecutive births and deaths of the links, 
$\Delta t_{\varsigma^2_i} = t^{(d_i)}_{\varsigma^2} - t^{(b_i)}_{\varsigma^2}$, where the index $i$ enumerates 
the birth ($b_i$) and the death ($d_i$) events, shown on Fig.~\ref{Figure3}A. First, we observe that the distribution of 
the link's effective lifetimes of both the two- and three-vertex connections has a bimodal shape. The relatively short 
($\Delta t \leq 10\tau$) lifetimes are exponentially distributed, implying that these connections are short-lived (the mode 
of the exponential distribution vanishes) and may be characterized by the effective decay times $\tau^{(2)}_{e}$ (links) 
that is about twice higher than $\tau$ and $\tau^{(3)}_{e}$ (triple connections) that is approximately equal to $\tau$ 
(Fig.~\ref{Figure2}). 
On the other hand, the bulging tails of the distributions shown on Fig.~\ref{Figure3}A,B represent an emergent population 
of long-lived pair and triple connections, i.e., a set of ``survivor'' simplexes that persist throughout almost the entire 
navigation period ($\Delta t_{\varsigma} \approx T_{tot}$). Thus, the net structure of the lifetimes' statistics suggests 
that the coactivity complex contains a stable ``core'' formed by a population of surviving simplexes, enveloped by a 
population of  rapidly recycling, ``fluttering,'' simplexes.

The \emph{mean} lifetime of each individual link, averaged over all the appearances across the entire navigation period, 
$\Delta t_{\varsigma_2} = \langle t^{(d_i)}_{\varsigma_2} - t^{(b_i)}_{\varsigma_2} \rangle_i$, can be approximated by 
a lognormal distribution with the mode $m_2 \approx 4$ mins (Fig.~\ref{Figure3}C), which corresponds to the mean lifetime 
of the ``fluttering'' connections (Fig.~\ref{Figure3}A). Similarly, a typical third-order simplex appears for about two mins 
(Fig.~\ref{Figure3}C), as suggested by the mean of the distribution shown on Fig.~\ref{Figure3}B. Thus, on average, both 
the coactivity graph $\mathcal{G}_{\tau}$ and the corresponding coactivity complex $\mathcal{F}_{\tau}$ exhibit 
persistent structures, despite rapid flickering of the individual connections.

The rejuvenation of simplexes also affects the frequency of their (dis)appearances. As shown on Fig.~\ref{Figure3}E,F, 
a typical link and a typical third order connection disappear about $4-5$ times during the navigation period, which is by 
an order of magnitude less than the links' activation rate (Fig.~\ref{Figure3}B). Thus, a typical simplex rejuvenates about 
$10$ times before getting a chance to decay. The histograms of the net lifetimes, i.e., of the total time that a given link 
or a clique spends in existence ($\Delta T_{\varsigma} = \Sigma_i \Delta t_{\varsigma,i}$) shown on Fig.~\ref{Figure3}G,H 
exhibit an even more salient contribution of the survivor simplexes. Note that the average net lifetime is approximately 
equal to the product of the mean effective lifetime and the mean number of appearances, $\Delta T_{e,\varsigma}\approx 
n_{e,\varsigma}\tau_{e,\varsigma}$, as expected.

\begin{figure}[!h]
\includegraphics[scale=0.83]{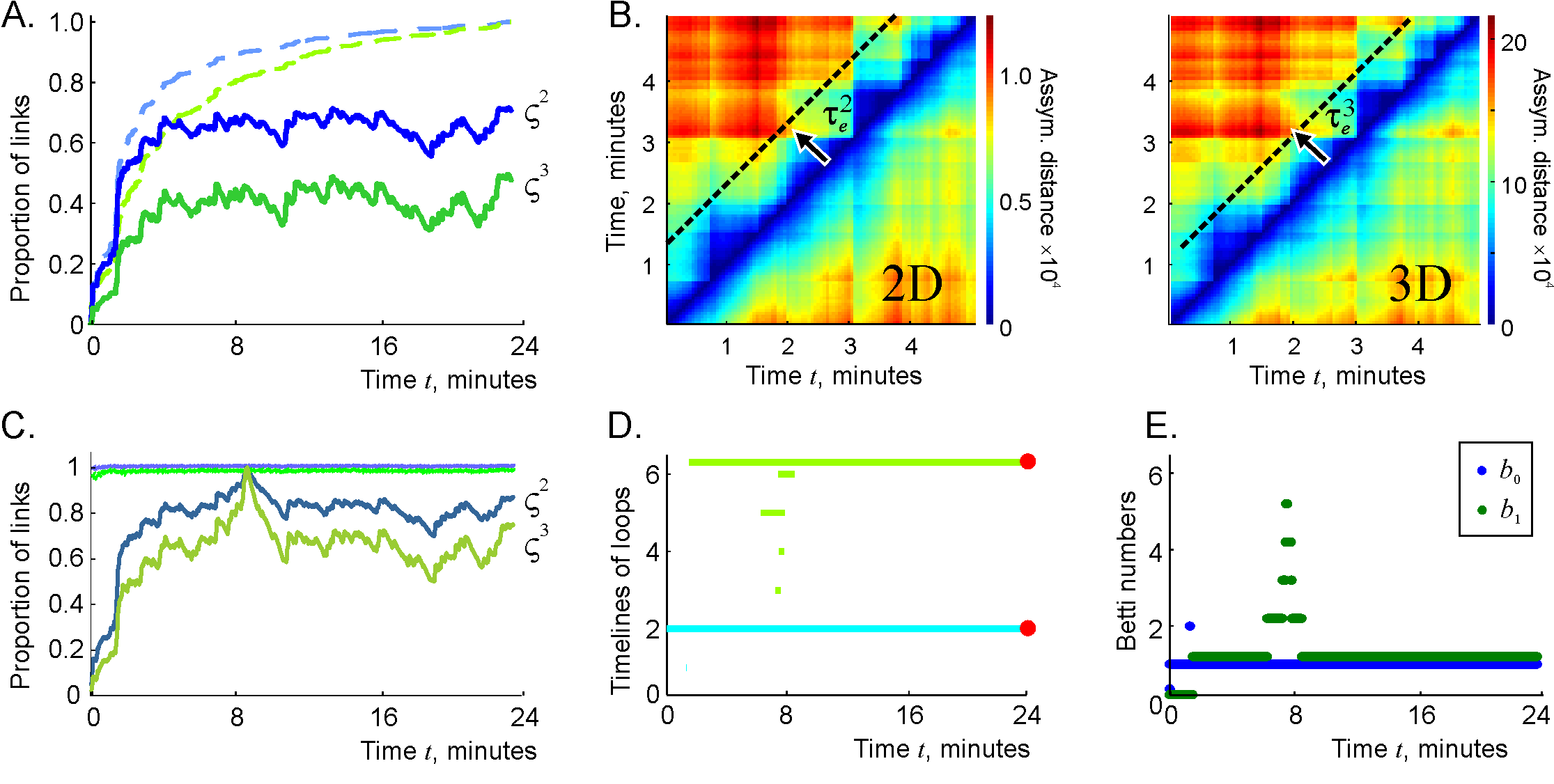}
\caption{{\footnotesize\textbf{Dynamics of the flickering complex.} A: The number of links $N_2(t)$ in the flickering complex 
$\mathcal{F}_{100}(t)$ (blue trace) compared to the number of links in the perennial complex $\mathcal{F}_{\infty}(t)$ (dashed 
light-blue trace). 
The corresponding numbers of triple connections $N_3(t)$ are shown by the green and the dashed light-green traces, respectively. 
B: The matrix of asymmetric distances $d^{(2)}_{ij}$ and $d^{(3)}_{ij}$ (note the difference in scales shown by the color bars), 
computed over a five minutes time interval. The changes in the coactivity complex accumulate at the $\tau_{e}$ timescale. 
C: The proportions of second and third order connections shared by the coactivity complexes at the consecutive moments, computed 
for links (top light-green line) and for the triple connections (top light-blue line) closely follow the $100\%$ mark, which implies 
that $\mathcal{F}_{\tau}(t)$ deforms slowly. The numbers of connections present at a moment of time when the coactivity complex 
is inflated ($t_{*} \approx 9$ min) that are also present at another moment $t$, $N_k(\mathcal{F}_{\tau}(t_{*})\cap\mathcal{F}_{\tau}(t))$, 
fluctuate around $\sim 80\%$ for links ($k = 2$, blue trace) and $\sim 60\%$ for triple connections ($k = 3$, green trace), implying 
that the same set of connections is being reactivated. D: Timelines of $0D$ (light-blue) and $1D$ (light-green) topological loops in 
$\mathcal{F}_{\tau}(t)$ indicate a splash of topological fluctuations near the inflation time $t = 9$ mins. During other periods, 
$\mathcal{F}_{\tau}(t)$ contains only one persistent loop in each dimension. 
E: The instantaneous Betti numbers, $b_0(\mathcal{F}_{\tau})$ and $b_1(\mathcal{F}_{\tau})$ increase around $t_{*} = 9$ min, 
but retain their physical values $b_0(\mathcal{F}_{\tau}) = b_1(\mathcal{F}_{\tau}) = 1$ for the rest of the navigation period, which 
implies that, despite flickering-induced deformations, the topological shape of the coactivity complex remains stable during almost 
the entire navigation period.}}
\label{Figure4}
\end{figure}

\textbf{Dynamics of the flickering coactivity complexes}. How does the decay of the connections affect the net structure 
of the flickering complex $\mathcal{F}_{\tau}$? As shown on Fig.~\ref{Figure4}A, the numbers of links $N_2(t)$ and of 
the triple connections $N_3(t)$ rapidly grow at the onset of the navigation and begin to saturate in about $t_s \approx 4$ 
mins (i.e., by the time when a typical link had time to make an appearance), reaching their respective asymptotic values 
in $t_a \approx 7$ mins. To put the size of the resulting flickering complex into a perspective, note that the number of 
simplexes in a decaying complex $\mathcal{F}_{\tau<\infty}$ can never exceed the number of simplexes that would have 
existed in absence of link decay, i.e., in the ``perennial'' coactivity complex, $\mathcal{F}_{\infty}\equiv \mathcal{T}$. 
Thus, the size of $\mathcal{F}_{\tau}$ can be characterized by the proportion of simplexes that happened to be actualized 
at that moment. As illustrated on Fig.~\ref{Figure4}A, these numbers fluctuate around $60\%$ for the second order simplexes 
and around $40\%$ for the third order simplexes, with the relative variances $\Delta N_2/N_2 \approx 12\%$ and 
$\Delta N_3/N_3 \approx 17\%$ respectively. In other words, the perennial coactivity complex $\mathcal{F}_{\infty}(t)$ 
loses about a half of its size due to the flickering of the simplexes, and fluctuates within about $15\%$ margins from the mean.

To quantify the changes in the complexes' structure as a function of time, we evaluated the number of two- and three-vertex 
simplexes that are present at a given moment of time $t_i$, but are missing at a later moment $t_j$, normalized by the size of 
$\mathcal{F}_{\tau}(t_i)$, i.e., $d^{(k)}_{ij} = N_k(\mathcal{F}_{\tau}(t_i)\setminus\mathcal{F}_{\tau}(t_j))/N_k(\mathcal{F}_{\tau}(t_i))$, 
$k = 2,3$. As shown on Fig.~\ref{Figure4}B, these numbers, which we refer to, respectively, as the second and third asymmetric 
distances between $\mathcal{F}_{\tau}(t_i)$ and $\mathcal{F}_{\tau}(t_j)$, rapidly grow as a function of temporal separation 
$|t_i - t_j|$. In fact, after a $\tau_{e}$-period, the difference between $\mathcal{F}_{\tau}(t_i)$ and $\mathcal{F}_{\tau}(t_j)$ 
becomes comparable to the sizes of either $\mathcal{F}_{\tau}(t_i)$ or $\mathcal{F}_{\tau}(t_j)$, which implies that the pool 
of simplexes in the simplicial complex is replenished at a $\tau_{e}$-timescale. However, the shape of the coactivity complex 
changes slowly: Fig.~\ref{Figure4}C demonstrates that nearly $100\%$ of the connections are shared at two consecutive moments, 
i.e., the changes in flickering complex from one moment of time to the next are marginal. As an additional illustration, we also 
computed the numbers of connections shared by the coactivity complex at a moment when $\mathcal{F}_{\tau}$ is inflated 
(about $t_{*} = 9$ mins, when a particularly large number of simplexes is simultaneously present) and other moments, when 
$\mathcal{F}_{\tau}$ is less bloated. The results shown on Fig.~\ref{Figure4}C indicate that the number of shared second and 
third order simplexes is respectively $N_2(\mathcal{F}_{\tau}(t_{*})\cap \mathcal{F}_{\tau}(t)) \approx 82\%$ and 
$N_3(\mathcal{F}_{\tau}(t_{*})\cap \mathcal{F}_{\tau}(t)) \approx 64\%$, i.e., that the recycling of the connections is due to 
partial reactivation of connections from the same pool.

Despite the rapid recycling of the individual simplexes, the large-scale topological characteristics of the flickering complex remain 
relatively stable. As demonstrated on Fig.~\ref{Figure4}D, after the initial stabilization period of about two minutes (which 
biologically may be interpreted as the initial learning period), $\mathcal{F}_{\tau}$ contains only one zero-dimensional and a 
single one-dimensional topological loop---as the simulated environment $\mathcal{E}$. Some topological fluctuations appear 
around $t \approx 9$ mins, as indicated by an outburst of short-lived spurious loops, most of which last for about a minute or 
less. After this period, the first two Betti numbers of $\mathcal{F}_{\tau}$ retain their physical values $b_0(\mathcal{F}_{\tau}) 
= b_1(\mathcal{F}_{\tau}) = 1$ (Fig.~\ref{Figure4}E). Since Zigzag homology theory allows tracing individual loops in 
$\mathcal{F}_{\tau}$ continuously across time, these persistent topological loops can be viewed as ongoing representations of 
the simply connected environment $\mathcal{E}$ and of the physical hole in it. Thus, the coactivity complex $\mathcal{F}_{\tau}$ 
preserves, for the most time, not only its approximate size, but also its topological shape---despite transience at the ``microscale'', 
i.e., at the individual simplex level. 

Physiologically, these results indicate that the large-scale topological information outlives the network's connections: 
although about a half of the functional links in the discussed case rewire within a $\tau_{e}$-period, the topological 
characteristics of the cognitive map encoded by the cell assembly network remain stable. In other words, a transient 
cell assembly network can encode stable topological characteristics of the ambient space, despite transience of the 
synaptic connections.

\begin{figure}[!h]
\includegraphics[scale=0.83]{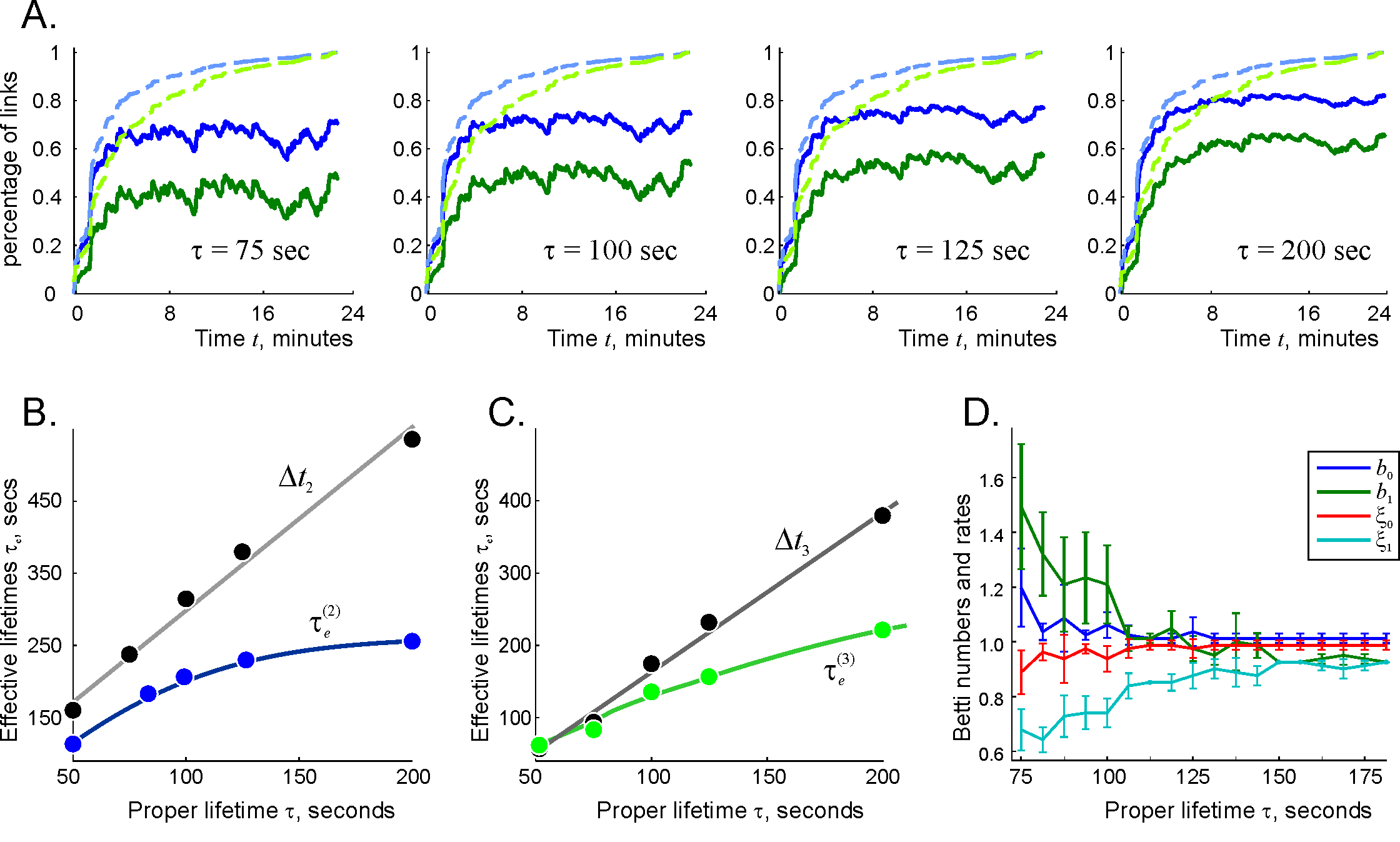}
\caption{{\footnotesize\textbf{$\tau$-dependence of the functional connections and of the topological loops.} 
A: The number of the two-vertex cliques ($N_2(\mathcal{F}_{\tau},t)$, green trace) and the number of three-vertex 
cliques ($N_3(\mathcal{F}_{\tau},t)$ blue trace) in the decaying complex, contrasted with the total number of two-vertex 
and three-vertex connections in the perennial complex across time ($N_2(\mathcal{F}_{\infty},t)$ is shown by dashed 
light-green line and $N_3(\mathcal{F}_{\infty},t)$ by light-blue line). The horizontal alignment of panels is used to 
emphasize increase in the asymptotic values $N_{2,3}(\mathcal{F}_{\tau},t\gg\tau)$.
B: The mean lifetimes of the ``fluttering'' links (i.e., the non-survivor links) in the coactivity complex are about twice longer 
than the proper link lifetimes, $\tau^{(2)}_{e} \approx 2\tau$ (blue line). The mean lifetimes of all the links in the coactivity 
complex (i.e., including the survivor, or ``core'' links) are about thrice longer than the proper lifetime, $\Delta t_{\varsigma_2} 
\approx 3\tau$ (gray line). 
C: Same dependences are shown for the triple connections. The effective lifetimes of the short-lived triple connections are 
approximately equal to proper link lifetime, $\tau^{(3)}_{e} \approx \tau$. 
D: The dependence of the Betti numbers of the flickering complex, $b_k(\mathcal{F}_{\tau})$, $k = 0,1$ and the 
corresponding percentages of the successful trials $\xi_k(\mathcal{F}_{\tau})$, on the proper decay time demonstrates 
that the topological fluctuations in $\mathcal{F}_{\tau}$ subside as $\tau$ increases.}}
\label{Figure5}
\end{figure}

\textbf{Dependence on the proper decay time $\tau$}. Next, we investigated the topological stability for a set of proper 
decay times $\tau$ ranging from one to five minutes. As expected, the number of simplexes in the flickering complex 
increases with growing $\tau$: in the studied map, the number of links raises from about $40\%$ at $\tau = 75$ 
secs to just under $60\%$ for $\tau = 200$ secs, whereas the number of the third order connections raises from 
$60\%$ to about $80\%$ (Fig.~\ref{Figure5}A). The distributions of the effective lifetimes for the short-lived (fluttering) 
connections retain their exponential shapes (Fig.~\ref{SFigure1}), 
with the means that are approximately proportional to the proper decay times, $\tau^{(2)}_{e} \approx 2\tau$ and 
$\tau^{(3)}_{e} \approx \tau$ (Fig.~\ref{Figure5}B,C). The contribution of the surviving simplexes also steadily grows 
with $\tau$ (Fig.~\ref{SFigure1}); 
as a result, the net average lifetimes, computed for the entire population of simplexes, grow faster
 $\Delta t_{\varsigma_2} \approx 3\tau$ and $\Delta t_{\varsigma_3} \approx 2\tau$.

As $\tau$ increases, the Betti numbers rapidly reduce to their physical values, $b_0(\mathcal{F}_{\tau}) = b_1(\mathcal{F}_{\tau}) = 1$: 
the lower is the connection decay rate, the smaller are the topological fluctuations generated in the flickering complex 
(Fig.~\ref{Figure5}D and Fig.~\ref{SFigure2}A,B). 
This is a natural result: since the ``perennial'' map $\mathcal{F}_{\infty}$ 
converges to a stable, topologically correct shape in a matter of minutes, the longer the simplexes survive in the decaying 
case ($\tau < \infty$) , the closer the topological barcode of $\mathcal{F}_{\tau}$ should be to the barcode of the environment 
$\mathcal{E}$. As shown on Fig.~\ref{Figure5}D, a stabilization of topological barcode is achieved around $\tau \sim 2$ mins. 
This value can also be naturally interpreted: for such $\tau$, the rat moving at the mean speed of about $25$ cm/sec has 
time to complete a circle around the central hole before a typical connection disappears, which allows the induced coactivity 
complex to capture this key feature of the environment and to contract the spurious topological loops. Note however, that 
this is only a qualitative argument since the expected lifetimes of over $63\%$ of links is smaller than $\tau$ and the lifetimes 
of $15\%$ of them live longer than $2\tau$.
 
\begin{figure}[!h]
\includegraphics[scale=0.83]{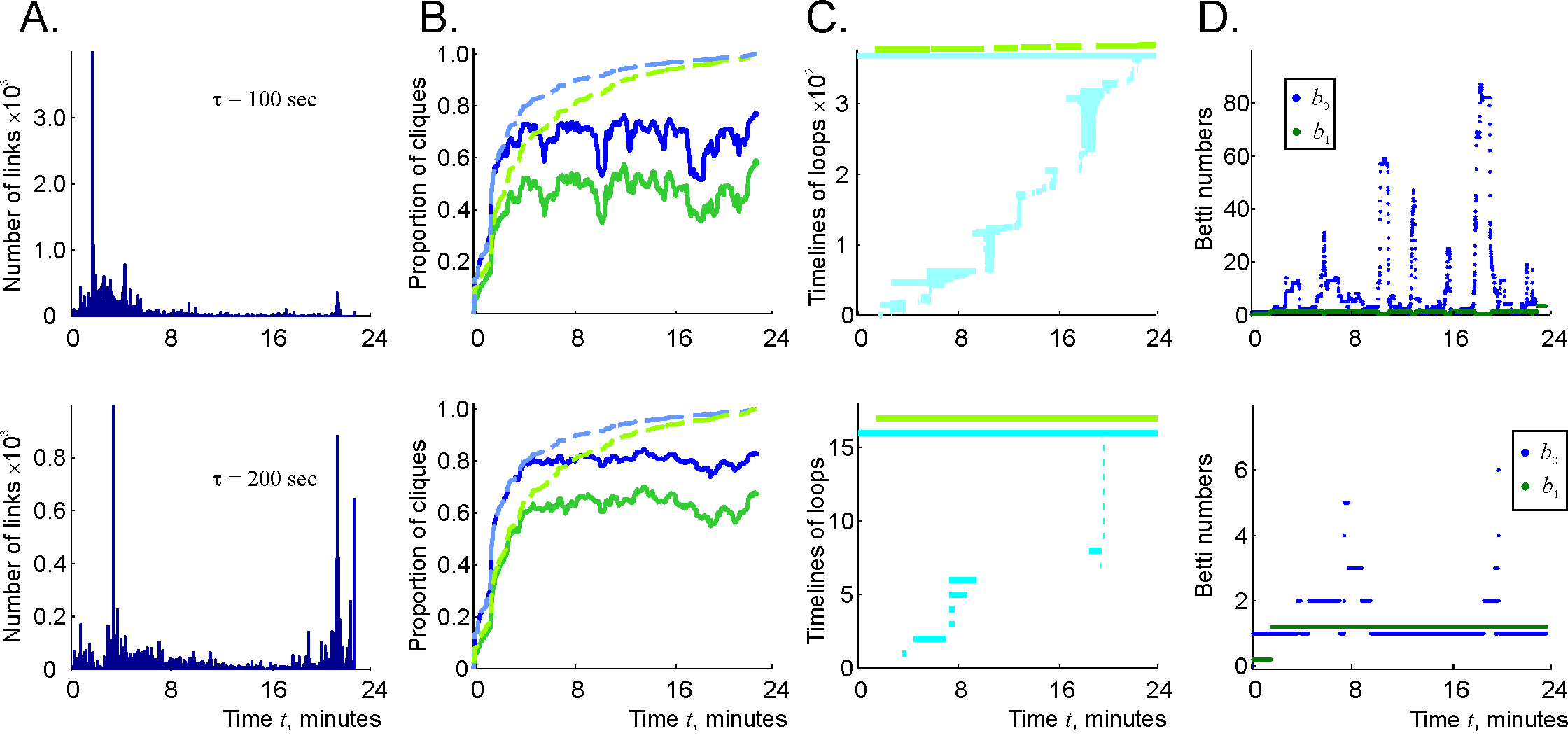}
\caption{{\footnotesize\textbf{Fixed connection lifetimes lead to topological instabilities.} A: The effective timelines of links 
with the proper decay time of $\tau = 100$ secs (top) and $\tau = 200$ secs (bottom). The contribution of the links retaining 
the original, singular proper decay time ($\tau_{e} = \tau$) is manifested in the sharp solitary peaks on the left sides of the 
histograms. The values to the left of that peak are produced by the ``boundary effect'': cutting the simulation at $T_{tot}$ 
produces timelines shorter than $\tau$. 
B: The distributions of the numbers of two- and three-vertex connections in (green and blue traces) vs. same numbers in 
the perennial complex $\mathcal{F}_{\infty}(t)$ (dashed lines) indicate that the number of instantiated connections in the 
case of the singular distribution (\ref{fix}) is higher than in the case of the distribution (\ref{decay}) (see Fig.~\ref{Figure4}A 
and Fig.~\ref{Figure5}A). 
As $\tau$ grows twofold (from 100 to 200 secs) the number of links $N_2(\mathcal{F}^{*}_{\tau})$ 
grows by $40\%$ and the number of triple connections $N_3(\mathcal{F}^{*}_{\tau}$) grows by $30\%$. 
C: The timelines of $0D$ (light-blue) and $1D$ (light-green) topological loops in the $\tau = 100$ secs (top) and in the 
$\tau = 200$ secs (bottom) case. The former produces hundreds of short-lived, spurious loops, while in the latter case there 
is about a dozen of loops that persist for about $50\%$ of the time. The behavior of the corresponding Betti numbers $b_0$ 
(blue) and $b_1$ (green) is shown on panel D.}}
\label{Figure6}
\end{figure}

\textbf{Fixed connection lifetimes}. To test how these results are affected by the spread of the link lifetimes, we investigated 
the case in which the lifetimes of all the links are fixed, i.e., the decay probability is defined by the function
\begin{equation}
p(t) = \begin{cases} 1 &\mbox{if } t = \tau \\ 
0 & \mbox{if } t \neq \tau. 
\end{cases}
\label{fix}
\end{equation}
while keeping the other parameters of the model unchanged. The results shown on Fig.~\ref{Figure6}A demonstrate that 
due to the rejuvenation effects, the range of the effective lifetimes widens and becomes qualitatively similar to the 
histograms induced by the decay distribution (\ref{fix}). As before, there appear two distinct populations of links: the 
short-lived links whose lifetimes concentrate around the singular proper lifetime $\tau$, and the ``survivor'' links, 
whose lifetimes approach $T_{tot}$.

However, the topological structure of the ``fixed-lifetime'' coactivity complex $\mathcal{F}^{*}_{\tau}$ differs dramatically 
from that of the decaying complex $\mathcal{F}_{\tau}$. As shown on Fig.~\ref{Figure6}C, $\mathcal{F}^{*}_{\tau}$ contains 
a large number of short-lived, spurious topological loops even for the values of $\tau$ that reliably produce physical Betti 
numbers in the case of the exponentially distributed link lifetimes. For example, at $\tau = 100$ secs, the zeroth Betti number 
of $\mathcal{F}^{*}_{\tau}$ hovers at the average value of $\langle b_0 \rangle \approx 40$, reaching at times $b_0 \sim100$, 
with nearly unchanged $b_1 = 1$, which indicates that at this decay rate, $\mathcal{F}^{*}_{\tau}$ brakes into a few dozens 
of disconnected, topologically contractible islets.

As the proper decay time increases, the population of survivor links grows and the disconnected pieces of 
$\mathcal{F}^{*}_{\tau}$ begin to pull together: at $\tau = 200$ secs, the Betti numbers $b_k(\mathcal{F}^{*}_{\tau})$ 
retain their  physical values for about a half of the time, yielding splashes of topological fluctuations during the other half 
(Fig.~\ref{Figure6}C,D). 

These differences between the topological properties of $\mathcal{F}_{\tau}$ and $\mathcal{F}^{*}_{\tau}$ indicate 
that the tail of the exponential distribution (\ref{decay}), i.e., the statistical presence of long-lasting connections is crucial 
for producing the correct topology of the flickering complex. Physiologically, this implies that the statistical spread of the 
connections' lifetimes plays important role: in absence of statistical variations, the dynamical cell assembly network fails 
to represent the topology of the environment.

\begin{figure}[!h]
\includegraphics[scale=0.83]{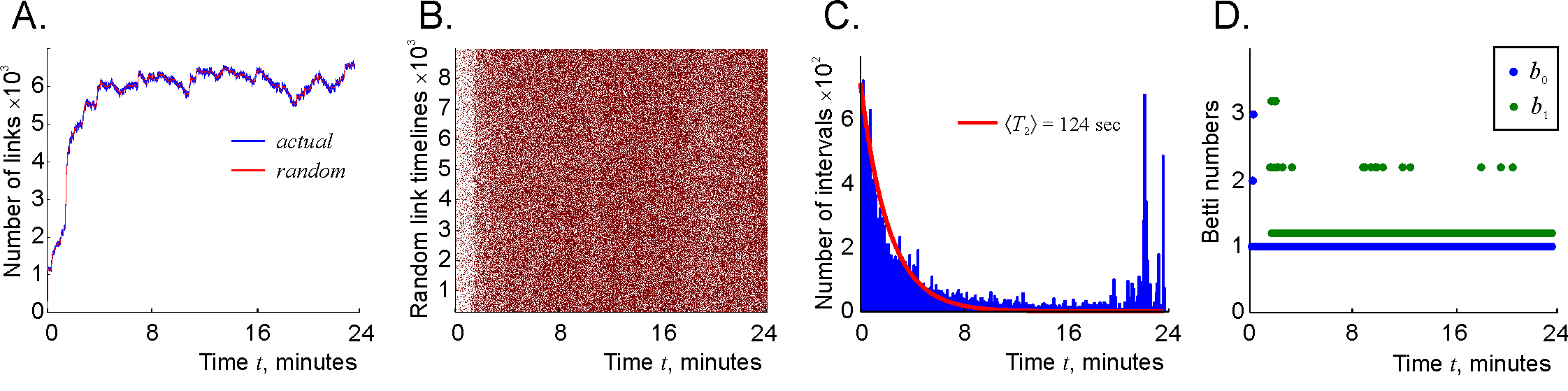}
\caption{{\footnotesize\textbf{Stochastic complex.} A: The number of links in the stochastic coactivity graph $\mathcal{G}_{r}(t)$ 
(blue trace) is the same as in the decaying coactivity graph $\mathcal{G}_{\tau}(t)$ (red trace). 
B: The links of the stochastic coactivity graph $\mathcal{G}_{r}(t)$ make instantaneous appearances and disappearances. 
Compare this chart to the timelines of the links in the decaying coactivity graph $\mathcal{G}_{\tau}(t)$ shown on 
Fig.~\ref{Figure2}B. 
C: Histogram of the link's net lifetimes in the stochastic graph indicates populations of short-lived and survivor links, 
similarly to the histogram shown on Fig.~\ref{Figure3}G. 
D: Betti numbers of the stochastic complex stabilize after the initial learning period of about four minutes, indicating 
the emergence of a stable topological shape of the simplicial complex with stochastically flickering simplexes.}}
\label{Figure7}
\end{figure}

\textbf{Randomly flickering connections}. These observations led us to another question: might the topology of the flickering 
complex be controlled by the shape of the lifetimes' distribution and the sheer number of links present at a given moment, 
rather than the specific timing of the links' appearance and disappearance? To test this hypothesis, we computed the number 
$n(t)$ of links in the decaying coactivity graph $\mathcal{G}_{\tau}(t)$ for $\tau = 100$ sec at every discrete moment of time 
$t$ (see Methods), and randomly selected the same number of links from the maximal available pool, i.e., from the graph 
$\mathcal{G}_{\infty}(t)$ that would have formed by that moment without links' decay (Fig.~\ref{Figure7}A). 
The collections of links randomly selected at consecutive moments of time can be viewed as instances of a stochastic 
connectivity graph $\mathcal{G}_{r}(t)$, i.e., as a graph whose links make random and instantaneous appearances and 
disappearances, in contrast with the decaying links of $\mathcal{G}_{\tau}(t)$ 
(compare Fig.~\ref{Figure7}B and Fig.~\ref{Figure2}B). 

Surprisingly, the random and the decaying graphs $\mathcal{G}_{r}(t)$ and $\mathcal{G}_{\tau}(t)$, as well as their 
respective clique complexes $\mathcal{F}_{r}(t)$ and $\mathcal{F}_{\tau}(t)$ possess a number of similar topological 
properties. First, the histogram of the \emph{net} lifetimes of the links in $\mathcal{G}_{r}(t)$ shown on Fig.~\ref{Figure7}C 
is bimodal, with an exponential part characterized by the mean $\langle T_2 \rangle = 124$ sec and a component 
representing a population of surviving connections, similar to the histograms shown on Fig.~\ref{Figure3}G,H. Second, the 
Betti numbers of the stochastic coactivity complex $\mathcal{F}_{r}$  converge to the Betti numbers of the environment 
in about $3$ mins---about as quickly as the Betti numbers of its decaying counterpart $\mathcal{F}_{\tau}$ (Fig.~\ref{Figure7}D). 
However, in contrast with $\mathcal{F}_{\tau}$, $\mathcal{F}_{r}$ keeps producing occasional fluctuations of one-dimensional 
spurious loops over the entire navigational period at a low rate (about $3\%$ of the time, Fig.~\ref{SFigure3}). 
Thus, the topological shapes of the random and the decaying coactivity complexes remain close. In other words, according 
to the model, the topological properties of the map encoded by a network with randomly formed and pruned 
connections are similar to the properties of a map produced by a network with decaying connections, as long 
as the net probability of the links' existence are same. In either case, rapidly rewiring connections do not preclude 
the appearance of a stable topological map.
 
\begin{figure}[!h]
\includegraphics[scale=0.83]{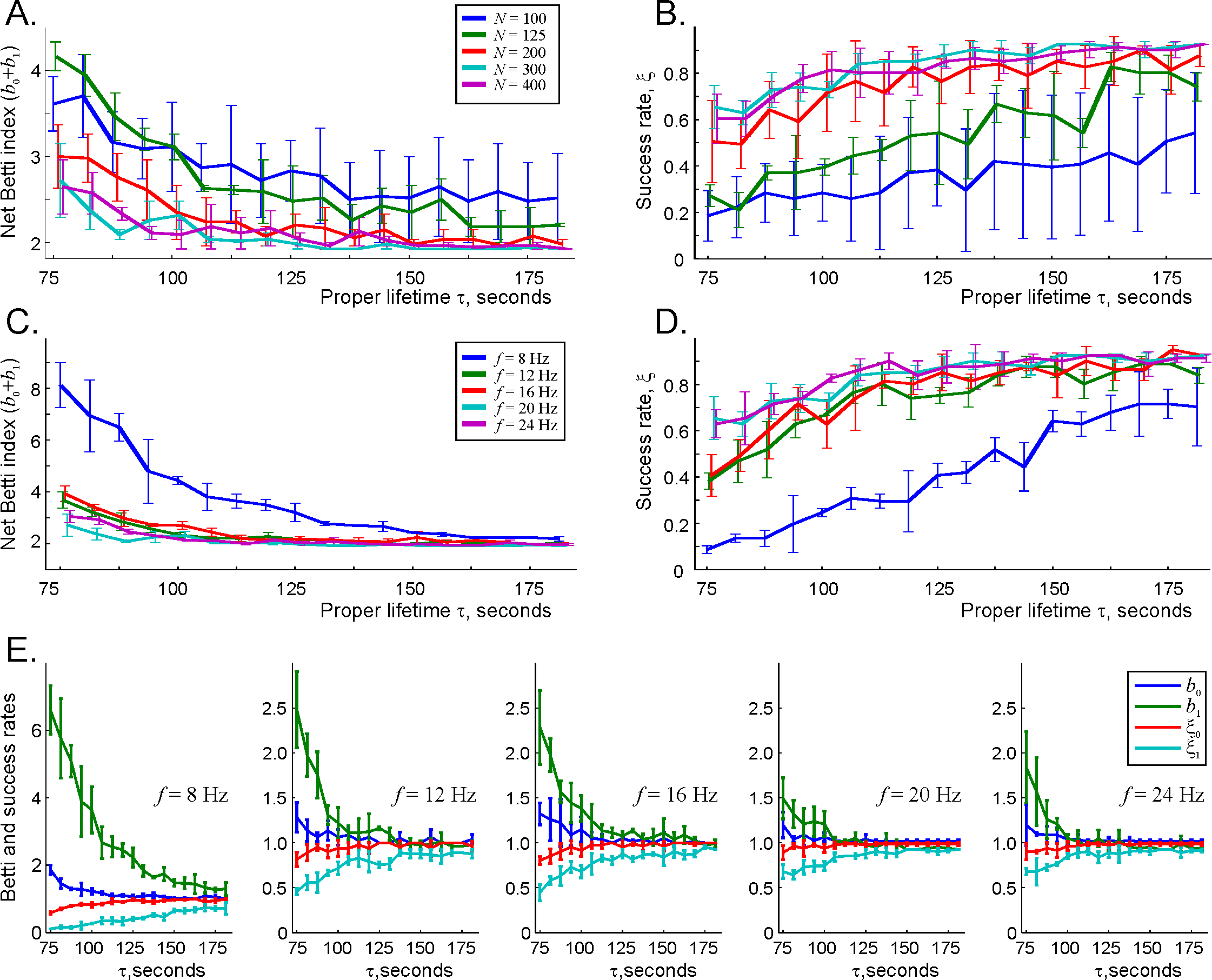}
\caption{{\footnotesize\textbf{Suppression of the topological fluctuations by increasing neuronal spiking activity.} 
A: As the number of active cells increases, the number of spurious topological loops drops. To compactify the information, 
we use the sum of the first two Betti numbers, $b = (b_0 + b_1)$, which describes the total number of $0D$ and $1D$ loops 
as a function of decay rate $\tau$, computed for several ensemble sizes. As the number of active cells with mean firing rate 
$f = 14$ Hz increases from $N = 100$ to $N = 400$ cells, the number of loops decrease from 3-4 (indicating at least one 
spurious loop in $0D$ or in $1D$) to the physical value $b_0(\mathcal{E})+b_1(\mathcal{E}) = 2$. 
B: The proportion of trials---the success rate, $\xi$---in which the coactivity complex produces the correct signature, $b_k(\mathcal{F}_{\tau})=b_k(\mathcal{E})$, as a function of the number of cells, $N$. Larger place cell ensembles 
tend to represents the topology of the environment more reliably. 
C: Sum of Betti numbers for different firing rates and $N$=300 cells. As the mean ensemble rate increases from $f = 8$ to 
$f = 24$ Hz, the spurious loops die out, i.e., the topological fluctuations in $\mathcal{F}_{\tau}$ are suppressed. 
D: The success rate $\xi_k$ as a function of the decay rate $\tau$, computed for a set of firing rates and $N = 300$. As before, 
the reliability of the map increases with the mean ensemble rate for the entire range of the proper decay times.
E: Betti numbers $b_k(\mathcal{F}_{\tau})$, $k = 0,1$, converge to their physical values $b_k(\mathcal{E})$ faster and their 
respective success rates $\xi_k$ grow more rapidly at higher firing rates.}}
\label{Figure8}
\end{figure}

\textbf{Compensatory mechanisms}. The turnover of memories (encoding new memories, integrating them into 
the existing frameworks, recycling old memories, consolidating the results, etc.) is based on adapting the synaptic 
connections across different brain regions \cite{OReilly}. In particular, these processes require a balanced contribution 
of both ``learning'' and ``forgetting'' components, i.e., forming and pruning the functional and/or physiological connections 
\cite{Kuhl, Murre}. The imbalances and pathological alterations in the corresponding synaptic mechanisms are observed in 
many neurodegenerative conditions, e.g., in the Alzheimer's disease, which is known to affect spatial cognition \cite{Selkoe}. 
However, interpreting the physiological meaning of these alterations is a highly nontrivial task, in particular because certain 
changes in neuronal activity may not be direct consequences of neurodegenerative pathologies.  For example, it is believed 
that neuronal ensembles may increase the spiking rates of the active neurons in order to compensate for the reduced synaptic 
efficacies \cite{Palop,Cacucci,Busche1,Busche2,Busche3,Minkeviciene,Siskova}. Such considerations motivate deep brain 
stimulation and other treatments that help to improve cognitive performance in animal models of Alzheimer's diseases and 
in Alzheimer's patients, by increasing the electrophysiological activity of hippocampal cells \cite{Laxton, Shirvalkar}.

Previous studies, carried out for the models of perennial cell assembly networks \cite{MB}, provided a certain theoretical 
justification for these approaches. For example, it was demonstrated that a place cell ensemble that fails to produce a reliable 
topological map of the environment due to an insufficient number of active neurons might be forced to produce a correct map 
by increasing the active place cells' firing rates. Similarly, the reduction in the firing rates or poor spatial selectivity of spiking 
may sometimes be compensated by increasing the number of active cells and so forth. Since the current model allows modeling 
networks with transient synaptic connections, we wondered whether it might point out other compensatory mechanisms, e.g., 
indicate a theoretical possibility to compensate for the reduced cell assemblies' lifetimes by increasing the spiking activity of the 
place cells. 

To that end, we varied the mean firing rate $f$ and the number of cells $N$ in the simulated place cell ensemble and 
studied the topological properties of the resulting coactivity complex as a function of the links' proper half-life, $\tau$. 
The results shown on Fig.~\ref{Figure8} demonstrate that indeed, increasing neuronal activity helps to suppress topological 
fluctuations in the flickering coactivity complex for a wide range of the connections' decay times. Moreover, these changes 
also increase the proportion $\xi$ of trials in which the place cell ensemble captures the correct signature of the environment.

Physiologically, these results indicate that recruiting additional active cells and/or boosting place cell firing rates 
allows to compensate for an overly rapid deterioration of synaptic connections, i.e., increasing neuronal activity 
stabilizes the topological map. In particular, a higher responsiveness of the Betti numbers of the flickering coactivity 
complex to an increase of the mean firing rate (Fig.~\ref{Figure8}C,E). 
as compared to the number of active place cells (Fig.~\ref{Figure8}A) suggests that targeting the active neurons' 
spiking may provide a better strategy for designing clinical stimulation methods.

\section{Discussion}
\label{section:discuss}

The formation and disbanding of dynamical place cell assemblies at the short- and intermediate-memory timescales 
enables rapid processing of the incoming information in the hippocampal network. Although many details of the 
underlying physiological mechanisms remain unknown, the schematic approach discussed above provides an 
instrument for exploring how the information provided by the individual cell assemblies combines into a large-scale 
spatial memory map and how this process depends on the physiological parameters of neuronal activity. In particular, 
the model demonstrates that a network with transient synaptic connections can successfully capture the topological 
characteristics of the environment.

Previously, we investigated this effect using an alternative model of transient cell assemblies, in which the connections
were constructed by identifying the pool of cells that spike within a certain ``coactivity window,'' $\varpi$, and building 
the coactivity graph $\mathcal{G}_{\varpi}$ by selecting the most frequently cofiring pairs of neurons \cite{MB}. The 
accumulation of topological information within each $\varpi$-period (physiologically, $\varpi$ can be viewed as the time 
over which the downstream networks integrate place cell outputs), was then described using persistent homology theory 
techniques. The results indicate that if $\varpi$ extends over 4-6 mins or more, the topological fluctuations in the 
flickering complex are suppressed and the topological shape of $\mathcal{F}_{\varpi}$ becomes equivalent to the shape 
of the environment. 

In the current model, enabled by the much more powerful Zigzag persistent homology theory \cite{Carlsson1,Carlsson2,Edelsbrunner}, 
we employ an alternative approach, in which the links of the coactivity graph appear instantly following pairwise place 
cell coactivity events. Thus, in contrast with the model discussed in \cite{MB}, the current model involves no accumulatoin 
of spiking information and no selection of the ``winning'' coactivity links, which one might hold responsible for stabilizing
the shapes of the flickering coactivity complexes. Nevertheless, the ``latency free'' model demonstrates the same effect: 
the large-scale topological shapes of resulting coactivity complexes stabilize, given that the connections decay sufficiently 
slowly and have sufficiently broadly distributed lifetimes. The connections' lifetimes required to achieve such stabilization 
are longer than in the input integration model ($\tau \approx 100$ sec vs. $\tau_{\varpi} \approx 10$ sec), which indicates 
that implementing a stable topological map in a physiological network with rapidly recycling functional connections may require 
integrating spiking information over an extended period and optimizing the network's architecture over this information. 
However, the fact that stable topological maps can emerge in different types of transient networks (including stochastically 
flickering networks) suggests that this is a generic effect that may explain appearance of stable cognitive representations 
of the environment in different physiological neuronal networks with ``plastic'' synaptic connections. In other words, the 
emergence of stable topological maps may represent a common ``umbrella'' phenomenon that can be implemented via 
different physiological mechanisms. 

In all cases, the model reveals three principal timescales of spatial information processing. First, the ongoing information 
about local spatial connectivity is rapidly processed at the working memory timescale, which physiologically corresponds 
to rapid forming and disbanding of the dynamical place cell assemblies in the hippocampal network. The large-scale 
characteristics of space as described by the instantaneous Betti numbers unfold at the intermediate memory timescale, 
whereas at the long-term memory timescale the topological fluctuations average out, yielding stable, qualitative 
information about the environment. Thus, the model reaffirms functional importance of the complementary learning systems 
for processing spatial information at different timescales and at different levels of spatial granularity \cite{OReilly,McClelland,Fusi}.

\section{Methods}
\label{section:methods}

\textbf{Place cell spiking activity} is modeled as a stationary temporal Poisson process with the maximal firing 
rate $f_c$ localized in a place field centered at $r_c$,
\begin{equation}
\lambda_c(r)=f_c e^{-\frac{(r-r_c)^2}{2s^2_c}}
\nonumber
\end{equation}
where $s_c$ defines the place field's size \cite{Barbieri}. An ensemble of $N$ place cells is therefore defined 
by $3N$ independent parameters, which we consider as random variables drawn from the stationary distributions, 
characterized by a mean firing rate, $f$, and the mean place field size $s$, i.e., by a point $(s, f, N)$ in a $3D$ 
parameter space. In addition, spiking is modulated by the $\theta$-oscillations---a basic cycle of the extracellular 
local field potential (LFP) in the hippocampus, with the frequency of $\sim 8$ Hz \cite{Buzsaki2}.
We study an ensemble of $N = 300$ place cells, with the typical maximal firing rate $f = 14$ Hz and the typical 
place field size $s = 20$ cm. 

\begin{figure}[!h]
\includegraphics[scale=0.83]{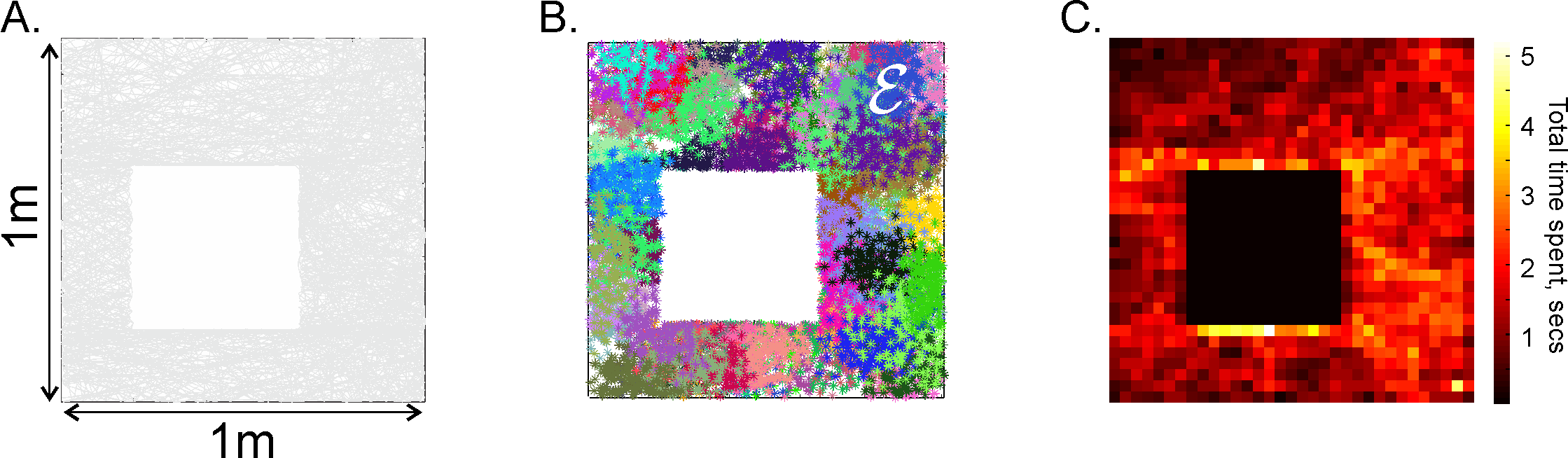}
\caption{{\footnotesize\textbf{Simulated Environment.} A: The trajectory covers a small planar arena $\mathcal{E}$ uniformly, 
without artificial circling or other ad hoc favoring of one segment of the environment over another. B: Simulated 
place field map $M_\mathcal{E}$. Clusters of dots of a particular color represent spikes produced by the corresponding 
place cells. C: A $2D$ histogram of the time spent by the animal in different locations---the occupancy map of $\mathcal{E}$.}}
\label{Figure9}
\end{figure}

\textbf{Place cell coactivity}. We consider a group of place cells $c_0, \ldots, c_k$ \emph{coactive}, if they produce 
spikes within two consecutive $\theta$-periods \cite{Arai,Mizuseki}. As a result, the time interval $[0, T_{tot}]$ splits 
into $1/4$ sec long time bins that define the discrete time steps $t_1,\ldots, t_n, \ldots$, used throughout the text 
\cite{Babichev1}.

\textbf{Simulated environment $\mathcal{E}$} represents a small $(1m \times 1m)$ square arena with a square 
hole in the middle, similar to the environments used in electrophysiological experiments \cite{Fu}. Fig.~\ref{Figure9}
shows the simulated trajectory, the layout of the place fields in $\mathcal{E}$--the place field map $M_\mathcal{E}$,
and the occupancy map. In \cite{Arai} we demonstrated that different parts of the environment can be learned independently 
from one another. Thus, knowing how learning works in small spatial domains, one can ``map out'' environments of 
different geometric and topological complexity. In particular, considering small environment shown above is sufficient 
for establishing the main effect (stability of global maps in transient networks) in general case.

\begin{figure}[!h]
\includegraphics[scale=0.83]{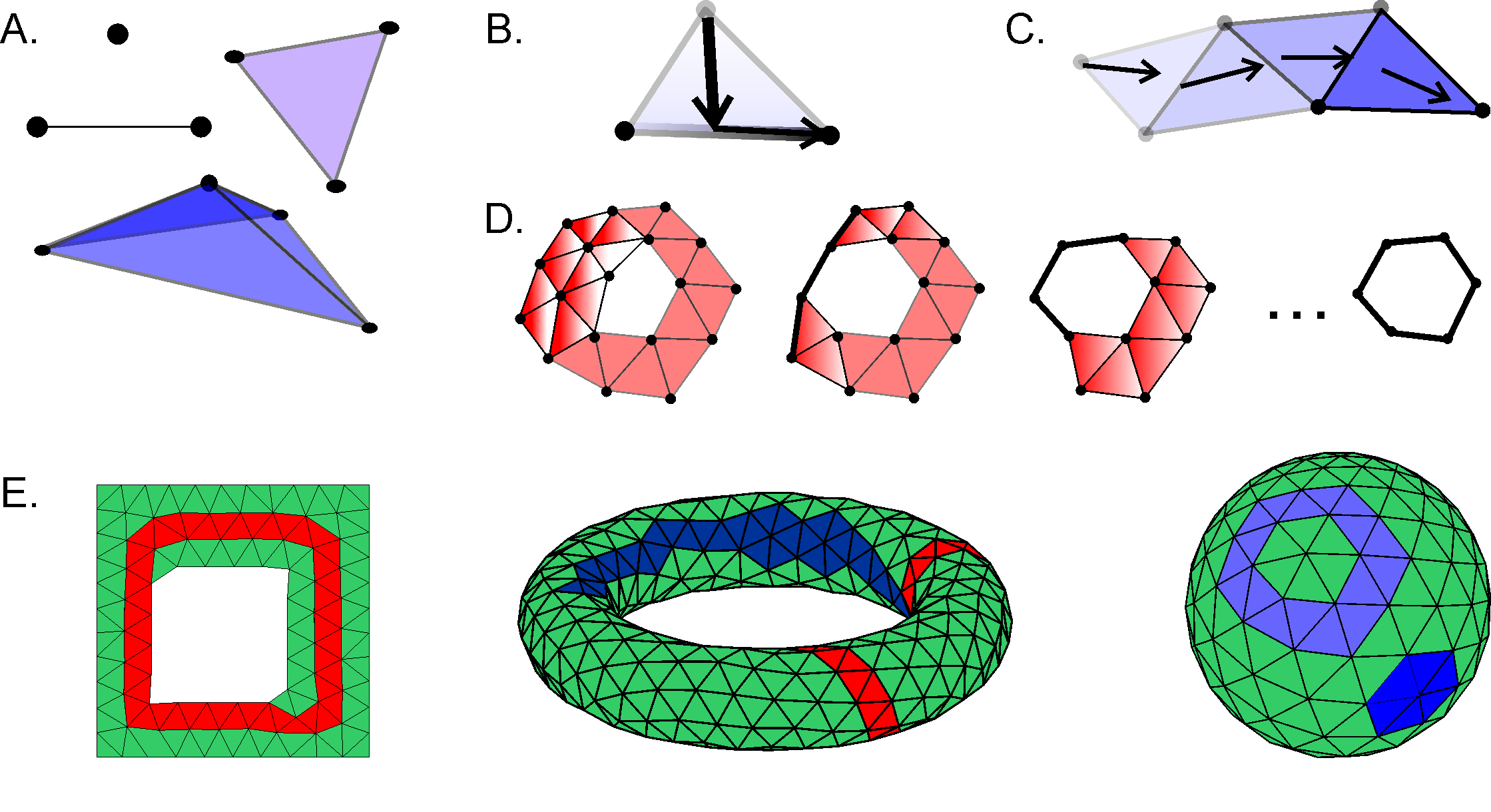}
\caption{{\footnotesize\textbf{Simplexes and simplicial complexes.} A: A zero-dimensional ($0D$) simplex $\sigma^0_i$ corresponds 
to a point vertex $v_i$; a one-dimensional ($1D$) simplex $\sigma^1_{ij}$---to a link between two vertexes $v_i$ and 
$v_j$; a two-dimensional ($2D$) simplex $\sigma^2_{ijk}$ ---to a filled triangle; a three-dimensional ($3D$) simplex 
$\sigma^3_{ijkl}$---to a filled tetrahedron, etc. The $n$ vertexes connected by the full set of $1D$ links form cliques, 
$\sigma^n$, of the corresponding order. B: A single simplex $\sigma^n$ is a contractible figure, i.e., it can be collapsed 
into one of its facets $\sigma^{n-1}$, then to a facet of lower dimensionality $\sigma^{n-2}$ and eventually to a point 
$\sigma^0$. Shown is a triangle contracting onto its bottom edge and then to the right vertex. 
C: A linear chain of simplexes bordering each other at a common face is also contractible. The shade of the triangles 
constituting the chain defines the order in which the triangles can be contracted (the lighter is the triangle, the sooner 
it contracts) and the arrows indicate the direction of the contractions. D: If a chain of simplexes loops onto itself and 
encircles a gap in the middle, then it is not contractible. Collapsing the triangles on the sides of such a closed chain 
produces an equivalent closed loop, which, ultimately, can be reduced to a non-contractible $1D$ loop, but not to a $0D$ 
vertex (the hole in the middle prevents that). Topologically, the deformed loops are equivalent to one another, i.e., they 
should be viewed as deformations of the same topological loop. E: Three simplicial complexes: a complex shaped as 
the environment $\mathcal{E}$ (see Fig.~\ref{Figure1}), a toroidal and a spherical complexes (figures obtained using 
MATLAB mesh generator \cite{Persson}). Non-contractible topological loops are shown as closed chains of red triangles 
and contractible loops are shown in shades of blue.}}
\label{Figure10}
\end{figure}

\begin{figure}[ht]
\includegraphics[scale=0.83]{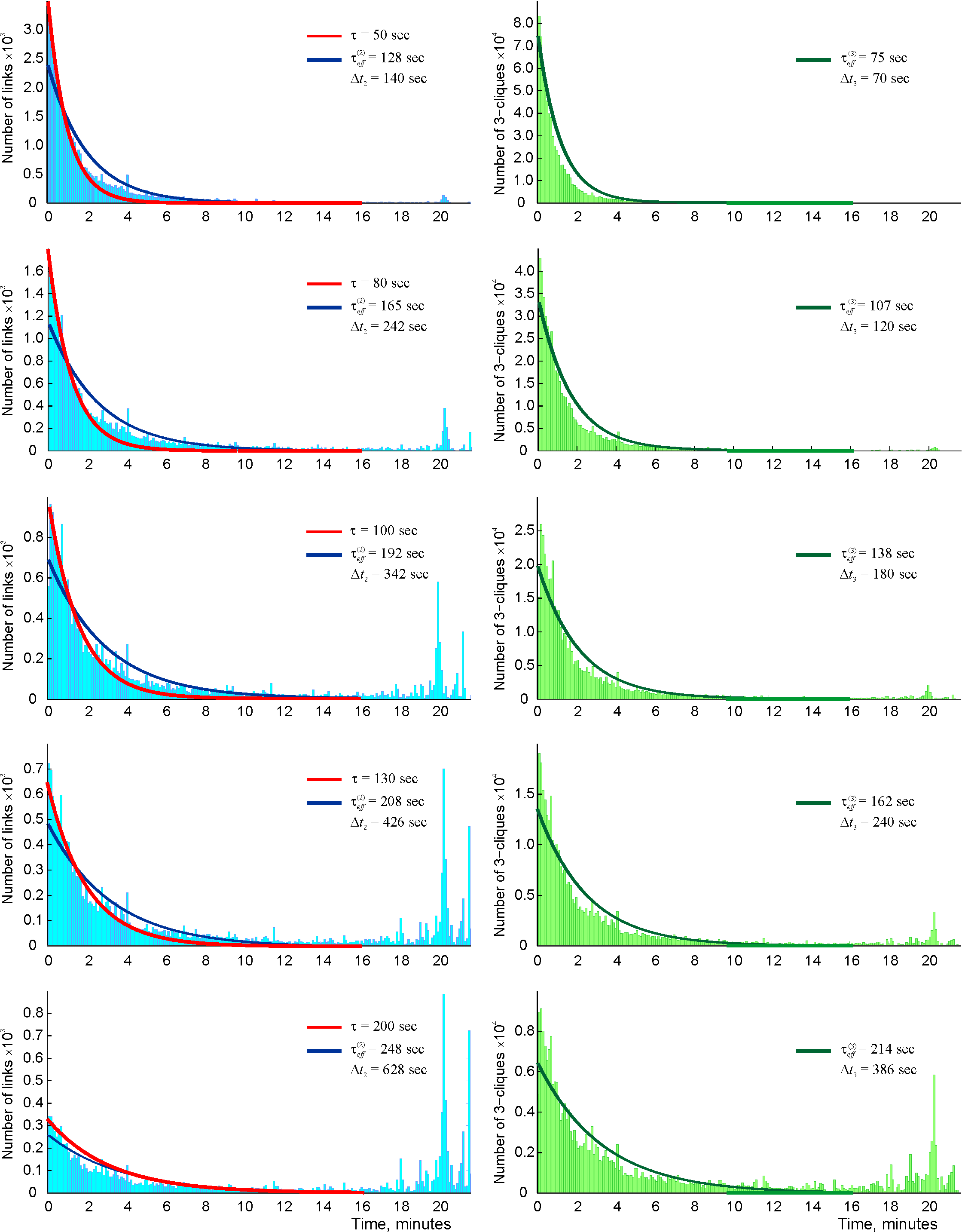}
\caption{{\footnotesize \textbf{Statistics of the connections' lifetimes.} A: Histograms of the intervals between consecutive 
births ($b$) and deaths ($d$) of the pairwise ($\Delta t_{\varsigma^2_i} = t^{(d_i)}_{\varsigma^2} - t^{(b_i)}_{\varsigma^2}$, 
left column of panels) and triple ($\Delta t_{\varsigma^3_i} = t^{(d_i)}_{\varsigma^3} - t^{(b_i)}_{\varsigma^3}$, right column 
of panels) connections, for five values of the proper decay times $\tau$. The red line outlines the exponentials with proper decay 
time $1/\tau$ and the dark-blue line shows the exponential fit of the histogram with the decay rate $1/\tau_{e}$, computed for the 
under 16 mins long intervals. The exponential fit to the histogram of the effective lifetimes of short-living triple connections 
($\Delta t_{\varsigma^3} < 10$ mins) is shown by dark-green line on the right panels. The mean lifetime for the entire population 
of links, $\Delta t_{k}$, is shown at the on each panel.}}
\label{SFigure1}
\end{figure}

\begin{figure}[ht]
\includegraphics[scale=0.8]{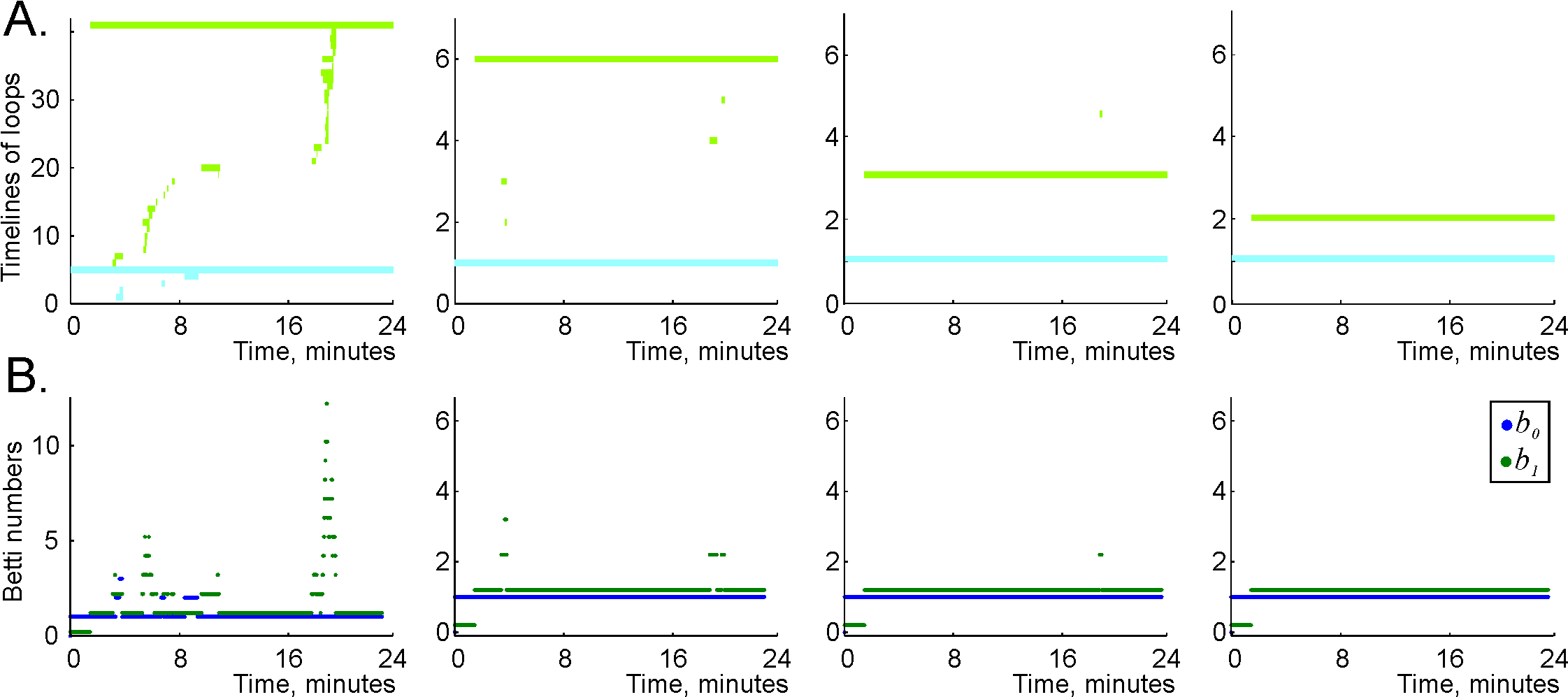}
\caption{{\footnotesize \textbf{Longer decay times suppress topological instabilities.} A: Timelines of $0D$ 
(light-blue) and $1D$ (light-green) topological loops in the flickering coactivity complex, computed for four values of the proper 
decay time $\tau$. B: The corresponding Betti numbers, $b_0(\mathcal{F}_{\tau})$ (blue) and $b_1(\mathcal{F}_{\tau})$ (green).}}
\label{SFigure2}
\end{figure}

\begin{figure}[ht]
\includegraphics[scale=0.83]{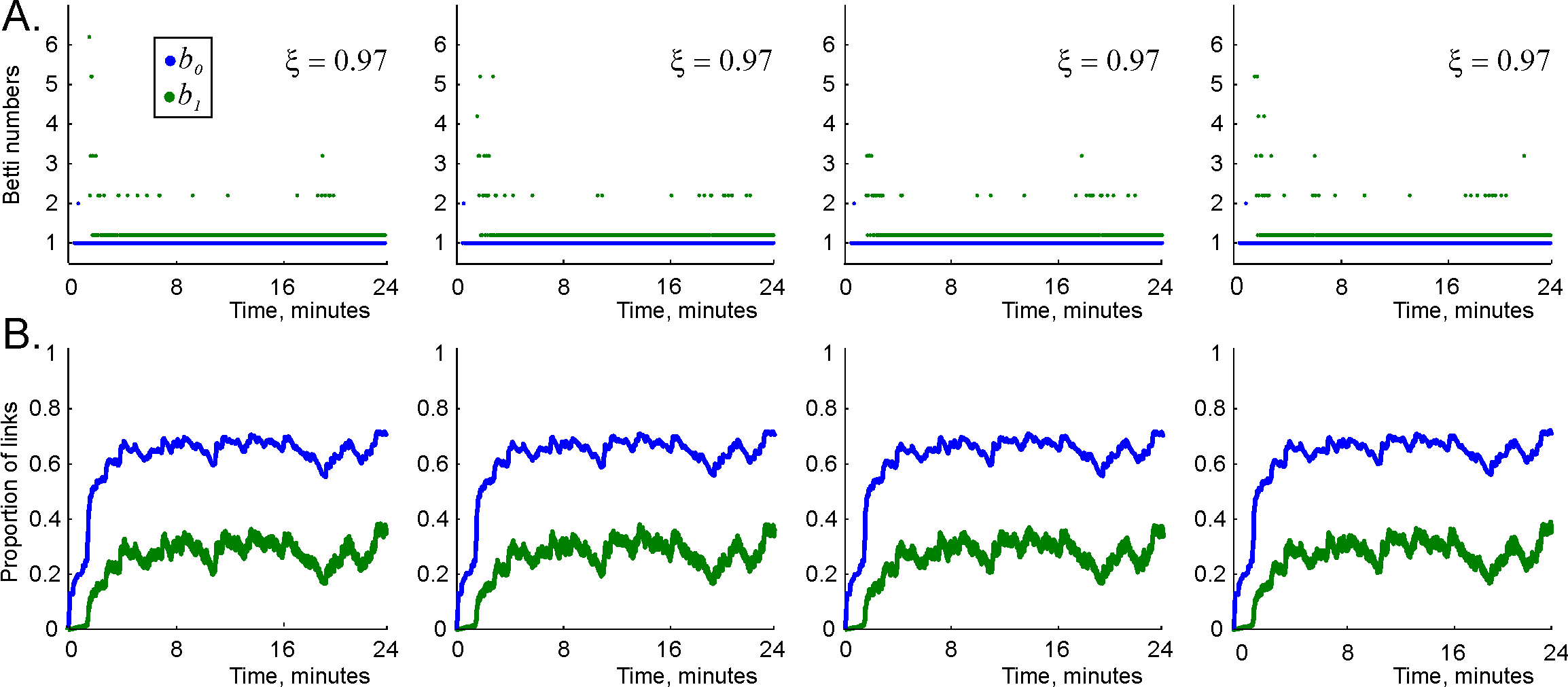}
\caption{{\footnotesize \textbf{Topological properties of the random complex.} A. Four tests of the 
topological behavior of the random complex $\mathcal{F}_r$ indicate that after initial period of about $3$ minutes, 
this complex produces occasional one-dimensional topological loops in only $3\%$ of the time (success rate $\xi = 0.97$ 
in all cases). B. The numbers of double and triple connection remains approximately the same from case to case.}}
\label{SFigure3}
\end{figure}

\textbf{Simplicial complexes.}
We use simplexes and simplicial complexes to represent combinatorially the topology of the neural activity. An abstract 
\emph{simplex} of dimensionality $n$ is a set containing $n+1$ elements. A subset of a simplex is  called its \emph{face}. 
A \emph{simplicial complex} is a collection of simplexes closed under the face relation: if a simplex belongs to a simplicial 
complex, then so do all of its faces (Fig.~\ref{Figure10}).

In the constructions studied in this paper, our simplicial complexes consist of coactive place cells. If all cells $\{c_0,\ldots,c_k\}$
are coactive within a given time window, then so is any subset of them, meaning coactive simplexes form a complex. In fact, 
because coactivity is defined for a pair of cells, our simplexes are precisely the cliques in the coactivity graph. 
A simplex $\{ c_0, \ldots, c_k \}$ is present if and only if all of its cells are pairwise coactive.

In flickering clique complexes, certain pairwise connections may decay over time, while others appear as time progresses. 
The effect on the simplicial complex is that some simplexes are removed from the complex, while others are added to it. 
So we get a sequence of ``flickering complexes,'' $X_i$, connected by alternating inclusions:
\[
X_1 \subseteq X_2 \subseteq X_3 \supseteq X_4 \subseteq X_5 \supseteq \ldots
\]
\textbf{Cycles, boundaries, and homology.}
A $k$-dimensional \emph{chain} is a set of $k$-dimensional simplexes (Fig.~\ref{Figure10}) that can be combined with 
suitable coefficients. If the coefficients form an algebraic field, then the chains form a vector space. Here we use the 
simplest algebraic field $\mathbb{Z}_2$, which consists of two Boolean values 0 and 1.  A boundary of the simplex is 
the sum of its one-dimension-lower faces:
$\bdry_k \{ c_0, \ldots, c_k \} = \sum_{i=0}^k \{ c_0, \ldots, c_{i-1}, c_{i+1}, \ldots, c_k \}$.
The map extends linearly to the entire simplicial complex, $X$, mapping its $k$-dimensional chains to its $(k-1)$-dimensional 
chains. The kernel of this map, i.e., all the chains without a boundary, is the set of \emph{cycles} of the complex, denoted by 
$\Zgr_k(X) = \ker \bdry_k$. The image of $\bdry_{k+1}$ consists of the $k$-dimensional chains that are \emph{boundaries} 
of some $(k+1)$-dimensional chains, denoted by $\Bgr_k(X) = \im \bdry_{k+1}$.

Cycles count ``$k$-dimensional holes'' in the complex. But not all such holes are independent of each other. We consider 
two cycles equivalent, or \emph{homologous}, if they differ by a boundary. Algebraically, one can verify that boundaries 
themselves have no boundaries, $\bdry_k \circ \bdry_{k+1} = 0$. In other words, boundaries are cycles.
This allows us to take a quotient, $\Hgr_k(X) = \Zgr_k(X) / \Bgr_k(X)$, called the $k$-dimensional \emph{homology} vector 
space. By definition, it considers two cycles equivalent, if their difference is a boundary of some $(k+1)$-dimensional chain. 
The dimension of this vector space, called the $k$-th Betti number, $\Betti_k(X) = \dim \Hgr_k(X)$, counts the number of 
independent holes in the topological space.

\textbf{Zigzag persistent homology.}
Given the sequence of flickering complexes above, we compute homology of each one. Inclusions between complexes induce 
maps between the homology vector spaces: the homology class of a cycle in the smaller complex maps to the homology class 
of the same cycle in the larger complex. Accordingly, we get a sequence of homology vector spaces, connected by linear maps:
\[
\Hgr_k(X_1) \to \Hgr_k(X_2) \to \Hgr_k(X_3) \ot \Hgr_k(X_4) \to \Hgr_k(X_5) \ot \ldots
\]
This sequence, called \emph{zigzag persistent homology}, generalizes ordinary persistent homology~\cite{Edelsbrunner}, where all 
the maps between homology groups point in the same direction. It is this generalization to the alternating maps that allows us 
to handle the flickering complexes.

On the surface, zigzag persistent homology tracks how the Betti numbers of the flickering complexes change. But the maps that 
connect homology vector spaces provide extra information. It is possible to select a basis for each vector space in this sequence, 
so that the bases for adjacent vector spaces are compatible~\cite{Carlsson1}. Specifically, we can select a collection of elements 
$\{ z_i^j \}_j$ for each vector space $\Hgr_k(X_i)$, such that the non-zero elements form a basis for the homology vector space 
$\Hgr_k(X_i)$ --- in other words, they represent a set of independent holes in $X_i$. Furthermore, such collections are compatible 
in the sense that adjacent basis elements map into each other: if we have a map $f: \Hgr_k(X_i) \to \Hgr_k(X_{i\pm1})$, then 
$f(z_i^j) = z_{i\pm1}^j$, if $z_i^j \neq 0$. The experiments in this paper use the algorithm of Carlsson et al.~\cite{Carlsson2} to 
compute such compatible bases.

It follows that the sequence of homology vector spaces can be decomposed into a barcode, where each bar represents the part 
of the sequence, where a particular basis element is non-zero. The bars capture when independent holes appear in the flickering 
complex, how long they persist, and when they eventually disappear.

\section{Acknowledgments}
\label{section:acknow}

The work was supported by the NSF 1422438 grant and by the Houston Bioinformatics Endowment Fund.

\newpage
\beginsupplement

\section{Supplementary Figures}

\begin{figure}[ht] 
\includegraphics[scale=0.83]{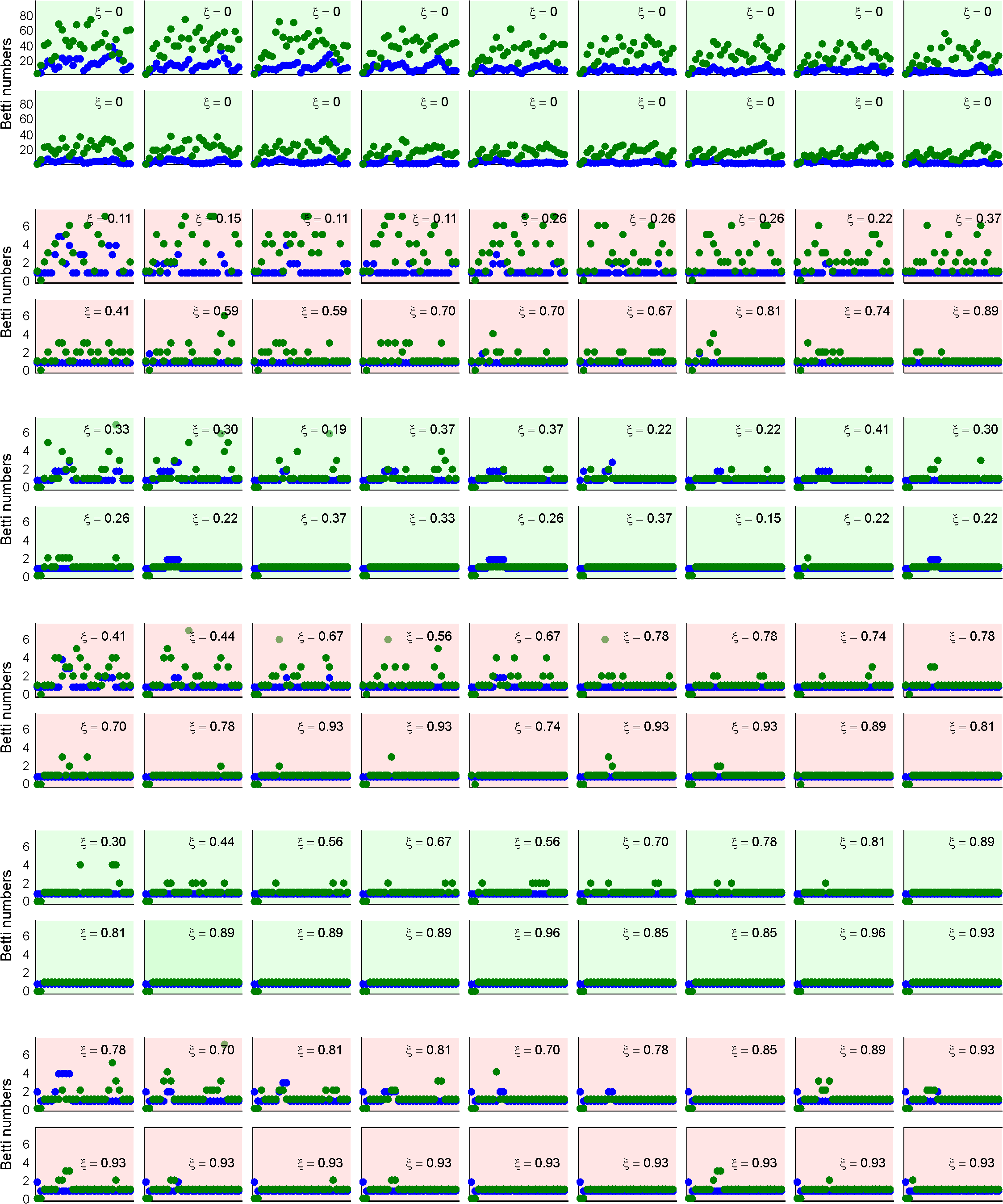}
\caption{{\footnotesize \textbf{Suppression of topological fluctuations by increasing place cell firing rates.} The six consecutive 
pairs of rows (colors alternate for illustrative purposes) correspond to the ensemble mean firing rate $f = 12$, 
$14$, $16$, $18$, $20$ and $24$ Hz. The proper decay time increases along each pair of rows from $\tau = 75$ 
to $\tau = 200$ secs, uniformly across the intermediate values. As $\tau$ increases, the percentage of times ($\xi$) 
during which the Betti numbers $b_k(\mathcal{F}_{\tau})$, $k = 0, 1$, remain equal to their physical values increases, 
for all ensemble mean firing rates. The higher is the ensemble mean frequency rate, the smaller are the topological 
fluctuations across the entire range of $\tau$s.}}
\label{SFigure4}
\end{figure} 

\end{document}